\documentclass{emulateapj}

\begin{document}

\title{The First Galaxies: Chemical Enrichment, Mixing, and Star Formation}

\shorttitle{The First Galaxies}

\author{Thomas H. Greif\altaffilmark{1}, Simon C. O. Glover\altaffilmark{2}, Volker Bromm\altaffilmark{3}, and Ralf S. Klessen\altaffilmark{2,4}}

\shortauthors{Greif et al.}

\affil{$^{1}$Max-Planck-Institut f\"{u}r Astrophysik, Karl-Schwarzschild-Stra\ss e 1, 85740 Garching bei M\"{u}nchen, Germany}
\affil{$^{2}$Zentrum f\"{u}r Astronomie der Universit\"{a}t Heidelberg, Institut f\"{u}r Theoretische Astrophysik, \\Albert-Ueberle-Stra\ss e 2, 69120 Heidelberg, Germany}
\affil{$^{3}$Department of Astronomy, University of Texas, Austin, TX 78712, USA}
\affil{$^{4}$Kavli Institute for Particle Astrophysics and Cosmology, Stanford University, Menlo Park, CA 94025, USA}

\email{tgreif@mpa-garching.mpg.de}

\begin{abstract}
Using three-dimensional cosmological simulations, we study the assembly process of one of the first galaxies, with a total mass of $\sim 10^8~M_{\odot}$, collapsing at $z\simeq 10$. Our main goal is to trace the transport of the heavy chemical elements produced and dispersed by a pair-instability supernova exploding in one of the minihalo progenitors. To this extent, we incorporate an efficient algorithm into our smoothed particle hydrodynamics code which approximately models turbulent mixing as a diffusion process. We study this mixing with and without the radiative feedback from Population~III (Pop~III) stars that subsequently form in neighboring minihalos. Our simulations allow us to constrain the initial conditions for second-generation star formation, within the first galaxy itself, and inside of minihalos that virialize after the supernova explosion. We find that most minihalos remain unscathed by ionizing radiation or the supernova remnant, while some are substantially photoheated and enriched to supercritical levels, likely resulting in the formation of low-mass Pop~III or even Population~II (Pop~II) stars. At the center of the newly formed galaxy, $\sim 10^5~M_{\odot}$ of cold, dense gas uniformly enriched to $\sim 10^{-3}~Z_{\odot}$ are in a state of collapse, suggesting that a cluster of Pop~II stars will form. The first galaxies, as may be detected by the {\it James Webb Space Telescope}, would therefore already contain stellar populations familiar from lower redshifts.\\
\end{abstract}


\section{Introduction}

One of the most important goals in modern astrophysics is to understand the end of the cosmic dark ages, when the first stars and galaxies transformed the simple early universe into a state of ever increasing complexity \citep{bl04a, cf05, glover05, bl07, bromm09}. The first stars, the so-called Population~III (Pop~III), were the source of hydrogen-ionizing UV photons, thus initiating the extended process of cosmic reionization \citep{fck06}. They also synthesized the first heavy chemical elements, beyond the hydrogen and helium produced in the big bang, to be dispersed into the pristine intergalactic medium (IGM) through supernova (SN) explosions and winds \citep[e.g.,][]{mfr01,wa08b}.

An intriguing possibility for the first stars is that some of them may have died as a pair-instability supernova (PISN), a peculiar fate predicted for progenitor masses in the range $140~M_{\odot}\la M_{*}\la 260~M_{\odot}$ \citep{brs67,hw02}. Current theory proposes a top-heavy initial mass function (IMF) for the first stars, with a characteristic mass $M_{*}\ga 100~M_{\odot}$ \citep{abn02,bcl02,on07,yoh08}. Together with the expectation that mass loss due to radiatively--driven winds is negligible at low metallicities \citep{kudritzki02}, one arrives at the robust expectation that at least a fraction of Pop~III stars should have died as PISNe. Compared to conventional core-collapse SNe, a PISN is distinguished by not leaving behind a compact remnant. Instead, the exploding star is completely disrupted, and {\it all} metals produced inside the Pop~III star are released into the surroundings, leading to metal yields of $y=M_{\rm Z}/M_{*}\sim 0.5$ \citep{hw02}. An abundant occurrence of PISNe in the early universe could thus have rapidly established a bedrock of metals, at least locally, of order $Z\ga 10^{-3}~Z_{\odot}$ \citep{greif07}. Recently, the extremely luminous SN~2007bi was observed in a nearby galaxy \citep{gal-yam09}, with characteristics, such as very large Ni masses, that seem to unambiguously point to a PISN origin. This discovery greatly strengthens the possibility for finding these events at high redshifts as well.

We here carry out cosmological simulations tracing the detailed assembly process of a primordial galaxy, taking into account the feedback effects from Pop~III star formation inside the minihalo progenitor systems. Our work extends the study by \citet{greif08}, which followed the virialization of gas in the galactic potential well under the idealized assumption of no such feedback. Specifically, we include radiative feedback, leading to the build-up of H~{\sc ii} regions around Pop~III stars \citep[e.g.,][]{wan04,abs06,yoshida07,greif09b}, as well as the mechanical and chemical feedback from a single PISN that explodes in the earliest minihalo progenitor. The latter feedback refers to the additional cooling that becomes available in metal-enriched gas, allowing the gas to reach lower temperatures, and to possibly fragment into low-mass Population~II (Pop~II) stars. It has been suggested that this transition in the star formation mode, from high-mass Pop~III to normal-IMF Pop~II, occurred once a minimum metal enrichment was in place, the so-called critical metallicity $Z_{\rm crit}\sim 10^{-6}-10^{-3.5}~Z_{\odot}$ \citep{bl03a,schneider06}. It is therefore important to identify any ``super-critical'' regions within a simulation box to distinguish between Pop~III and Pop~II star formation sites. The overall goal is to predict the properties of the first galaxies, to be observed with the {\it James Webb Space Telescope (JWST)}, planned for launch in $2014$ \citep{gardner06}. One key ingredient necessary for such predictions is to quantify the amount and distribution of metals inside the emerging galaxy, just prior to the onset of its initial starburst. This would allow us to constrain the detailed mix of stellar populations, and consequently to arrive at observables such as broad--band colors, emission line signatures, and luminosities \citep[e.g.,][]{johnson09}.

A complementary empirical probe of the first stars is given by ``Stellar Archaeology'', the study of abundance patterns in metal-poor Galactic halo stars \citep{bc05,fjb07,fjb09}, and of globular clusters in the Milky Way and other galaxies \citep[e.g.,][]{helmi06,pkg06}. To infer the characteristics of the first SNe, and therefore to constrain the primordial IMF from the observed fossil chemical record, we need to better understand the in situ physical conditions in the regions where the first low-mass stars formed. This again requires us to simulate the transport of metals from the first SNe in a realistic cosmological setting. A related problem is the apparent absence of a PISN abundance pattern in any of the known metal-poor stars \citep{tumlinson06}. It has been argued that such a PISN signature may be hidden in stars with relatively high metallicities, $Z\sim 10^{-2.5}~Z_{\odot}$, due to the extremely high yield from even a single explosion \citep{kjb08}. Our simulations can assess the viability of this scenario with much improved physical realism.

We trace the mixing and dispersal of metals, originating in a PISN inside a Pop~III minihalo at $z\simeq 30$, all the way to their reassembly into the growing potential well of the first galaxy at $z\simeq 10$. The physics governing this transport is highly complex. Some aspects, such as the advection of metals during hierarchical structure formation, is reliably modeled by our Lagrangian smoothed-particle hydrodynamics (SPH) approach. Other processes, such as hydrodynamical instabilities and turbulence, are not well resolved in our simulation. We may therefore miss the fragmentation of gas within the dense shell of the SN remnant, which could result in the formation of an intermediate generation of stars \citep{mbh03,on08,whalen08b,nho09}. However, we have developed an efficient algorithm to approximately model the mixing that results from these sub-grid effects as a diffusion process \citep{greif09a}. The corresponding diffusion coefficient is evaluated assuming a simple mixing-length approach to turbulent transport even in the supersonic regime \citep{kl03}, with  quantities that are entirely local. This allows us to greatly improve on our earlier, ``ballistic'' transport of discrete metal packets \citep{greif07}, and consequently to derive meaningful metallicity distribution functions.

The structure of our work is as follows. In Section~2, we describe our numerical methodology and the procedure to set up the initial conditions. We then discuss the assembly process of the first galaxy, and the resulting distribution of metals within our simulation volume (Section~3). In Section~4, we investigate the gas properties in the different sites where second-generation star formation could occur. We conclude by summarizing and assessing the cosmological implications of our results. All distances quoted in this paper are in proper units, unless noted otherwise.

\section{Numerical methodology}

\subsection{Simulation setup}
We perform our simulations in a cosmological box of size $1~{\rm Mpc}^{3}$ (comoving), initialized at $z=99$ with a fluctuation power spectrum corresponding to a concordance $\Lambda$ cold dark matter ($\Lambda$CDM) cosmology with matter density $\Omega_{m}=1-\Omega_{\Lambda}=0.3$, baryon density $\Omega_{b}=0.04$, Hubble parameter $h=H_{0}/100~{\rm km}~{\rm s}^{-1}~{\rm Mpc}^{-1}=0.7$, where $H_{0}$ is the present Hubble expansion rate, spectral index $n_{s}=1.0$, and a normalization $\sigma_{8}=0.9$ \citep{spergel03}. We take the chemical evolution of the gas into account by following the abundances of H, H$^{+}$, H$^{-}$, H$_{2}$, H$_{2}^{+}$, He, He$^{+}$, He$^{++}$, and $e^{-}$, as well as the three deuterium species D, D$^{+}$, and HD.  We include all relevant cooling mechanisms, i.e., H and He collisional ionisation, excitation and recombination cooling, bremsstrahlung, inverse Compton cooling, and collisional excitation cooling via H$_{2}$ and HD \citep{gj07}. We explicitly include H$_{2}$ cooling via collisions with protons and electrons, which is important for the chemical and thermal evolution of gas with a significant fractional ionization \citep{ga08}.

In a preliminary run with $64^3$ gas and dark matter (DM) particles, we locate the formation site of the first $\simeq 10^{8}~M_{\odot}$ halo. This object is massive enough to cool through Ly$\alpha$ emission and fulfil our prescription for a ``first galaxy'' \citep{greif08}. We subsequently carry out a standard hierarchical zoom-in procedure to achieve high mass resolution in the comoving volume of the galaxy \citep[e.g.,][]{nw94,tbw97,gao05}. We apply three consecutive levels of refinement, such that a single parent particle is replaced by a maximum of $512$ child particles. The particle masses for DM and gas in the highest resolution region containing the comoving volume of the galaxy with an extent of $\simeq 300~{\rm kpc}$ (comoving) are $\simeq 33~M_{\odot}$ and $\simeq 5~M_{\odot}$, respectively. The mass resolution is $\simeq 400~M_{\odot}$, so that the Jeans mass at a density of $n_{\rm H}=100~{\rm cm}^{-3}$ and temperatures near the cooling threshold set by the cosmic microwave background (CMB) is well resolved.

\subsection{Feedback by the first star}

The first minihalo collapses at a redshift $z\simeq 30$ near the location of the nascent galaxy. Once its central density exceeds $n_{\rm H}=100~{\rm cm}^{-3}$, we assume that a single Pop~III star with $200~M_{\odot}$ forms, near the middle of plausible values determined in numerical studies \citep{on07}. However, for the purpose of determining the properties of the ensuing H~{\sc ii} region, we use the ionization, dissociation, and heating rates for a $50~M_{\odot}$ Pop~III star \citep{schaerer02}, which limits the severity of the radiative feedback. We note that the SN remnant will likely propagate somewhat further in this case, since pressure equilibrium with the ambient medium will be achieved at later times \citep[see also][]{greif07}. On the other hand, the photoheating from nearby stars may mitigate this effect and lead to only minor deviations from the true metal distribution. The corresponding surface temperature, luminosity, and stellar lifetime may be found in \citet{schaerer02}, while details of the ray-tracing algorithm are described in \citet{greif09b}. Two key assumptions of this model are that the gas is optically thin to molecule-dissociating radiation, and that the effects of stellar evolution are negligible, which would otherwise lead to a decrease in the number of ionizing photons emitted at the end of the main sequence \citep{marigo01,schaerer02}.

At the end of its brief lifetime, the progenitor is predicted to explode as a PISN, ejecting of order $100~M_{\odot}$ in metals and $10^{52}~{\rm erg}$ of kinetic energy \citep{hw02}. In the simulation, we assign the above energy and metal content to the central $200~M_{\odot}$ of gas as described in \citet{greif07}. The explosion is initialized at the end of the free expansion phase, where the momentum of the swept-up mass leads to a transition to the Sedov--Taylor phase. This simplified treatment is roughly valid as long as the central density is low, which is established by the previous photoheating from the star \citep[e.g.,][]{jgb07}. All particles within the initial ejecta are assigned a metal yield of $50~\%$, consisting of C, O, and Si and relative abundances corresponding to a solar composition. Although this is somewhat inconsistent with the results of \citet{hw02}, who show that the actual C and Si abundances differ roughly by a factor of four from the solar values, the oxygen abundance is reproduced correctly. Since this is the most common element, we expect this caveat to not qualitatively affect our conclusions. We note that other elements such as Mg, S, Ca, and Fe are produced as well, but are negligible in terms of providing additional cooling \citep{ss06a}.

\subsection{Mixing of metals}

A key difficulty in modeling chemical mixing in SPH simulations is that no inherent mass flux between particles exists. Most simulations simply work around this caveat \citep[e.g.,][]{scannapieco05b,ksw07}, while others implement ad hoc methods to smooth the metals within the kernel \citep[e.g.,][]{wiersma09}. A more realistic treatment clearly requires a physical model for the distribution of metals between particles. We have recently developed a diffusion-based algorithm to accomplish this task \citep{greif09a}. On all scales of interest, chemical mixing proceeds via turbulent motions that cascade down to smaller and smaller scales. By quantifying the degree of turbulence on the smallest resolved scale, one can estimate the local degree of mixing. The exchange of metals between particles is then modeled as a diffusion process, with the diffusion coefficient set by the smoothing length and the velocity dispersion within the kernel. The SPH equations for diffusion are well known \citep[see][]{monaghan05}, and only minor modifications are required to apply this machinery to chemical mixing. We note that a similar method has recently been used in the context of IGM enrichment by massive galaxies \citep{sws09}.

\subsection{Metal cooling}

As a consequence of a more accurate metallicity estimate, we can model the additional cooling provided by heavy elements. For this purpose we assume that C, O, and Si are produced with solar relative abundances. This is a relatively good approximation for the cooling provided by the heavy elements ejected by a $200~M_{\odot}$ PISN. For these species, our model distinguishes between two temperature regimes. At temperatures below $2\times 10^{4}~{\rm K}$, we use a slightly modified version of the chemical network presented in \citet{gj07}. This treatment follows the chemistry of C, C$^{+}$, O, O$^{+}$, Si, Si$^{+}$, and Si$^{++}$, in addition to the primordial species mentioned above, and accurately models the effects of fine structure cooling from C, C$^{+}$, O, Si, and Si$^{+}$. It does not include the effects of cooling from molecules such as CO or H$_{2}$O, which is unimportant at the densities probed in this study \citep{gj07}. A modification to the \citet{gj07} treatment is the use of more accurate rate coefficients for the collisional excitation of the fine-structure lines of C and O by atomic hydrogen, taken from the recent calculations of \citet{akd07}. The second, more important modification is that we keep C and Si in their first ionized states, since they are easily ionized by a soft UV background, which is likely to be present at $z\la 20$ \citep[e.g.,][]{bl03a}.

At temperatures above $2\times 10^{4}~{\rm K}$, a full non-equilibrium treatment of metal chemistry, such as we use in cold gas, becomes significantly more costly, owing to the increasing number of ionization states that become available. We therefore adopt a cooling rate derived from the values given in \citet{sd93} for gas in collisional ionization equilibrium. \citet{sd93} give total cooling rates (including hydrogen and helium line cooling, and bremsstrahlung) for metallicities $Z=0$, $10^{-3}$, $10^{-2.0}$, $10^{-1.5}$, $10^{-1.0}$, $10^{-0.5}$, $1.0$, and $10^{0.5}~Z_{\odot}$. By subtracting off the values for the zero metallicity case from the other cases, we can isolate the contribution of the metals. For metallicities in the range $10^{-3}<Z<10^{0.5}~Z_{\odot}$, we can then determine the metal cooling rate by interpolation of these values. For metallicities $Z<10^{-3}~Z_{\odot}$ and $Z>10^{0.5}~Z_{\odot}$, we must extrapolate from these values, which is inevitably more inaccurate, although we note that in the former case metal cooling is ineffective compared to H and He cooling.

In view of the significant uncertainties that exist regarding the nature and the quantity of the dust produced by Pop~III SNe \citep{sfs04,bs07,nozawa07,cd09}, we do not include dust in our present simulation. We do not expect this to lead to a significant error: although dust cooling may be extremely important during the late stages of protostellar collapse in low metallicity gas \citep[see, e.g.,][]{omukai05,cgk08}, it is not an effective coolant at the gas densities resolved in this study.

\subsection{Feedback by second-generation stars}

The thermal and chemical evolution of the IGM is complicated by the fact that minihalos in the vicinity of the SN remnant collapse and form stars, which photoheat the gas that is eventually incorporated into the galaxy. Additional feedback by accretion from stellar remnant black holes or subsequent SN explosions may become important as well. However, in the present work, we choose to not include more than one SN, since we believe it is essential to understand the effects of a single SN before attempting to follow the evolution of multiple, interacting SNe. We therefore assume that all stars formed in minihalos collapse directly to black holes after they die. Recent investigations have shown that accretion onto these stellar remnant black holes is inefficient as a result of the low density of the gas after photoheating \citep{jb07,awa09}, at least for regions of the universe that are not part of an unusually large overdensity \citep[e.g.,][]{gao05}. This leads us to neglect radiative feedback from early miniquasars \citep[e.g.,][]{km05b}.

In light of the complexity induced by multiple star formation sites, we have chosen to run two distinct simulations: in simulation~A (Sim~A), we insert a $50~M_{\odot}$ Pop~III star whenever the density in a minihalo exceeds $n_{\rm H}=100~{\rm cm}^{-3}$, and subsequently solve for the H~{\sc ii} region according to the prescription in Section~2.2. After of order $50$ stars have formed, we stop the ray-tracing routine and instead form sink particles. This ensures that the simulation does not become exceedingly complex. Once a sink particle is created, all gas particles within the smoothing length corresponding to the above density threshold are immediately accreted. Further accretion is governed by the criterion that a particle must fall within this smoothing length (details on the implementation may be found in \citet{jappsen05}). In simulation~B (Sim~B), we {\it always} form sink particles and do not include any radiative feedback beyond that of the very first star. This distinction allows us to investigate the effects of photoheating on the distribution of metals and the assembly of the galaxy.

\section{Pre-galactic enrichment}
In this section, we discuss the evolution of the SN remnant and its influence on the surrounding medium, followed by a detailed investigation of the distribution of metals in both simulations. This will then allow us to discuss the consequences of metal enrichment for second-generation star formation in Section~4.

\subsection{Assembly of the first galaxy}

The expansion of the SN remnant into the IGM is depicted in Figures~1 and 2, where we show the central $100~{\rm kpc}$ (comoving) of Sim~A and B. After $\simeq 15~{\rm Myr}$, the SN remnant has propagated to $\simeq 1~{\rm kpc}$ in radius, roughly the size of the original H~{\sc ii} region. Both simulations are identical at this point, since no other minihalos have collapsed yet. In a few cases, the shock ahead of the SN remnant has disrupted neighboring minihalos that are just in the process of collapsing. However, most minihalos remain unaffected since they are either too far away or have collapsed to high enough densities \citep{cr08}. The distinction between disrupted and largely unaffected minihalos leads to an interesting result that will be discussed in Section~4.

After roughly $100~{\rm Myr}$, or a Hubble time at $z\simeq 30$, the SN remnant comes into pressure equilibrium with the surrounding medium and stalls. During this period, multiple neighboring minihalos undergo runaway collapse and form stars. In Sim~A, the ensuing H~{\sc ii} regions then partially overlap with the SN remnant and the comoving volume of the nascent galaxy. At this point, the complexity of the simulation dramatically increases and a multi-phase medium forms. In Figure~3, we show the distribution of the gas in density and temperature space within the high resolution region at $z\simeq 10$, excluding gas accreted onto sink particles. It occupies more than six orders of magnitude in density, ranging from $n_{\rm H}=4\times 10^{-5}$ to $100~{\rm cm}^{-3}$, and more than three orders of magnitude in temperature, ranging from $T=10$ to $4\times 10^{4}~{\rm K}$. Most of the gas is shock-heated to the virial temperature and cools once the density exceeds $n_{\rm H}\sim 1~{\rm cm}^{-3}$, but some parcels of gas in Sim~A are photoheated to $10^{4}~{\rm K}$ and have cooled down to $T\simeq 1000~{\rm K}$ by $z\simeq 10$. This warm, diffuse medium resides in the low-density IGM at $n_{\rm H}\sim 10^{-3}$~--~$10^{-4}~{\rm cm}^{-3}$ and will not be available for further star formation for $\ga 10^8$~yr, the approximate IGM free-fall time. A third phase, consisting of gas shock-heated by the SN remnant, has cooled down from an initial temperature of $\sim 10^{8}~{\rm K}$ to $\sim 10^{4}~{\rm K}$ and has become indistinguishable from the relic H~{\sc ii} region gas.

\begin{figure*}
\begin{center}
\resizebox{17cm}{20.9cm}
{\unitlength1cm
\begin{picture}(17,20.9)
\put(0.0,19.4){\includegraphics[width=17.0cm,height=1.5cm]{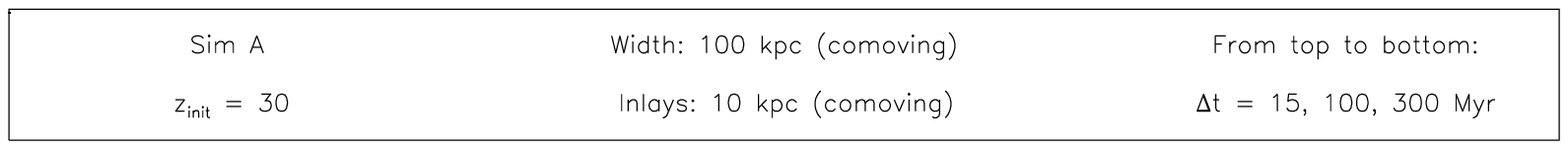}}
\put(0.0,13.7){\includegraphics[width=5.6cm,height=5.6cm]{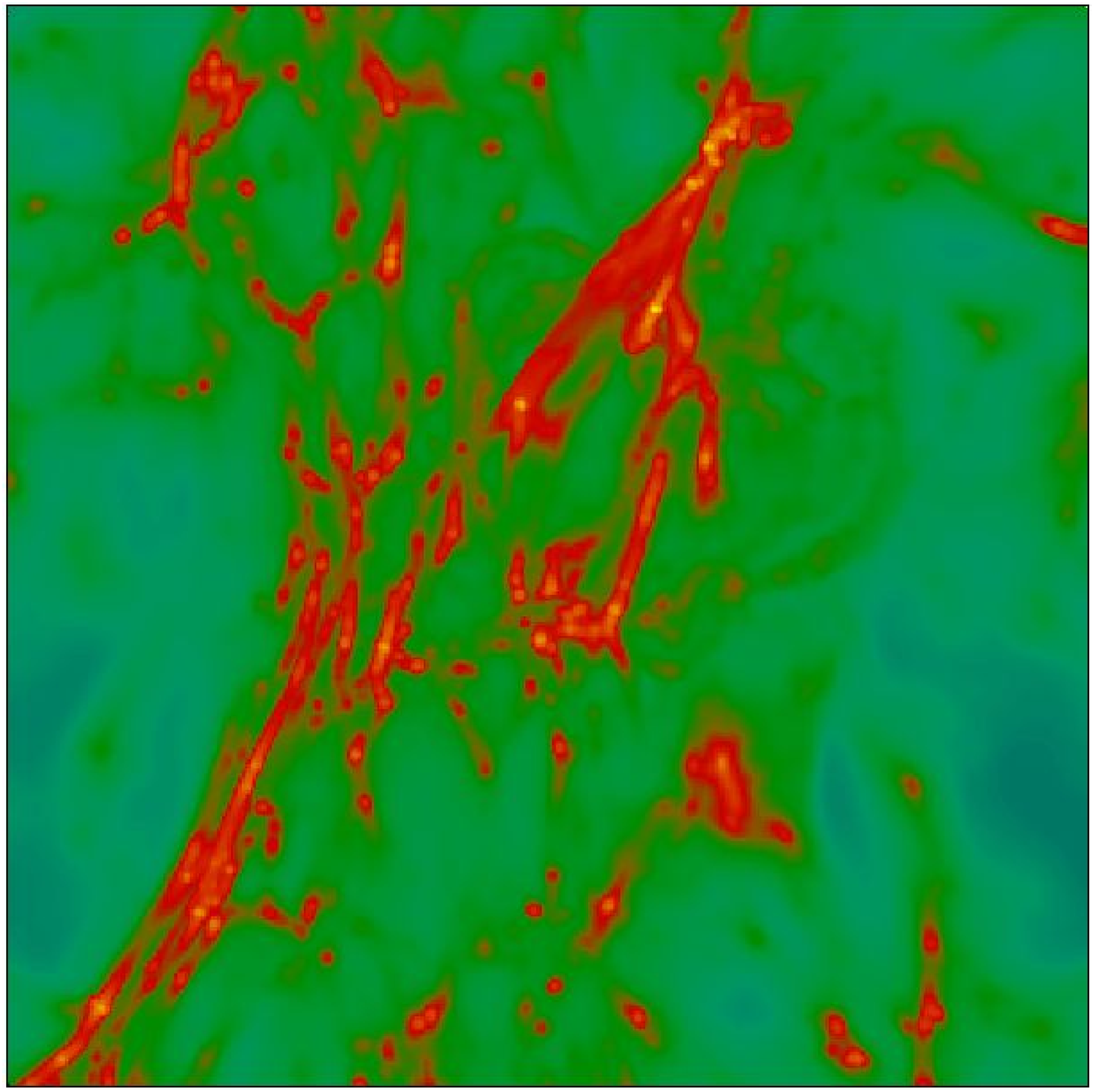}}
\put(5.7,13.7){\includegraphics[width=5.6cm,height=5.6cm]{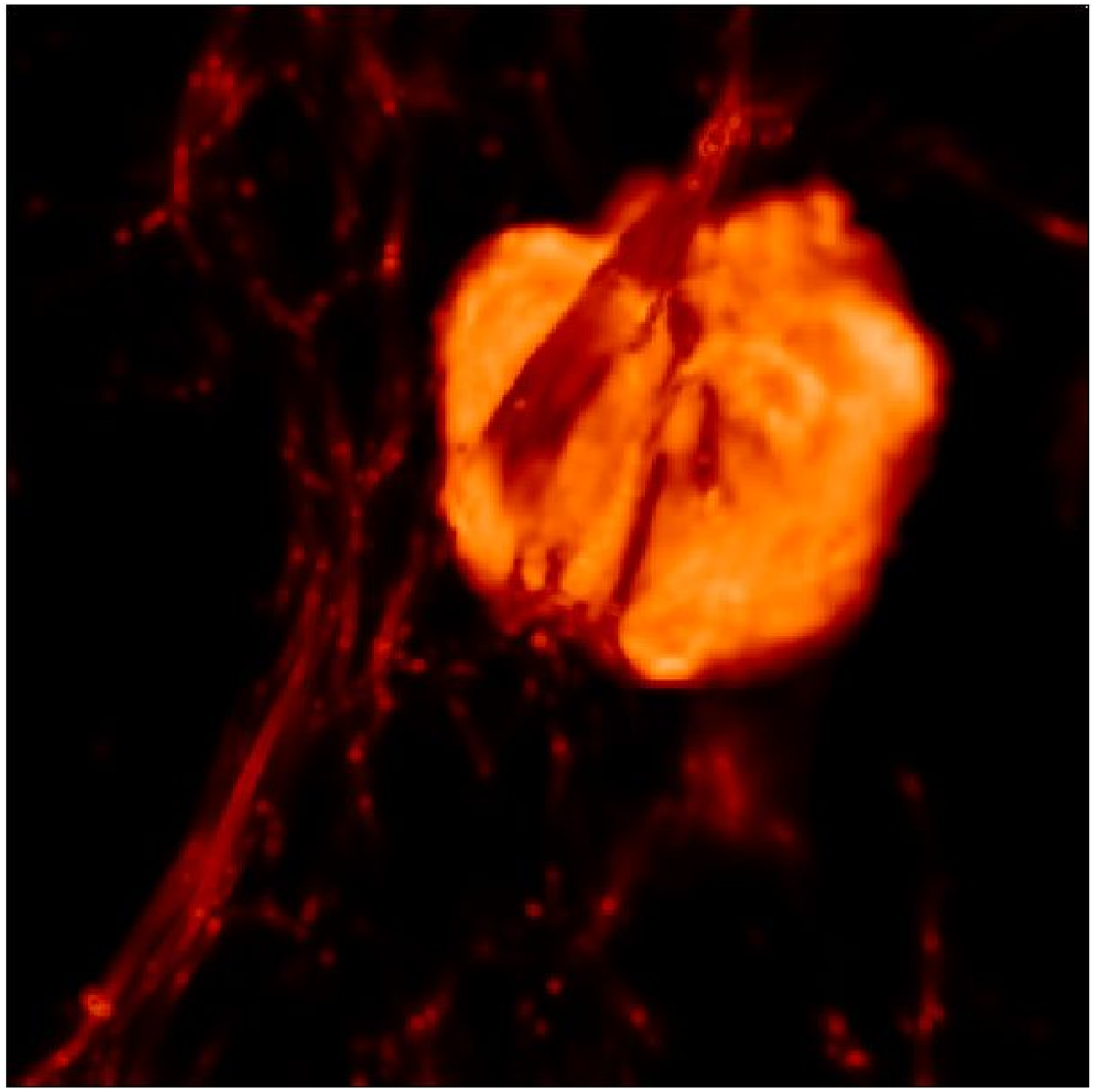}}
\put(11.4,13.7){\includegraphics[width=5.6cm,height=5.6cm]{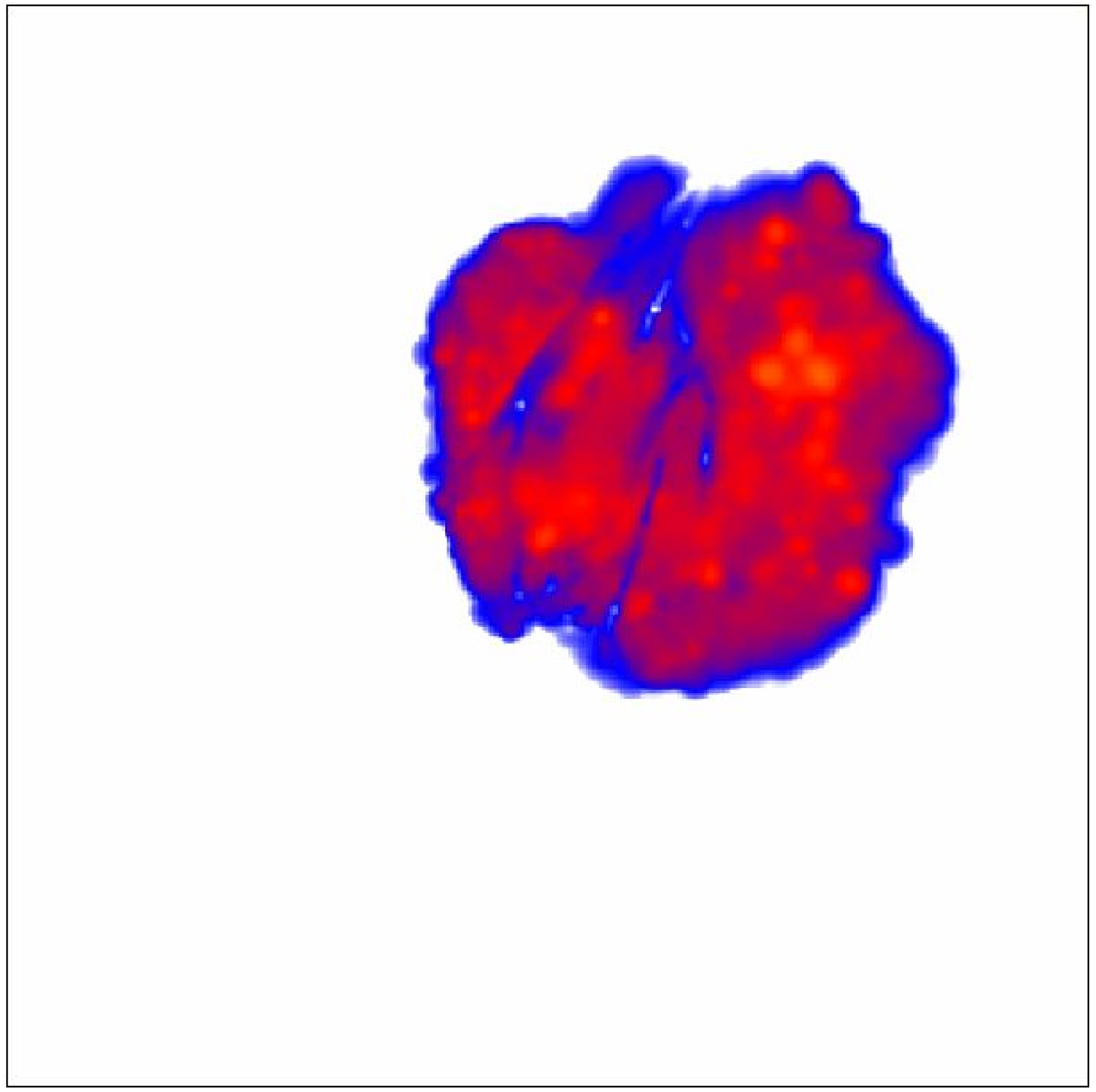}}
\put(0.0,7.9){\includegraphics[width=5.6cm,height=5.6cm]{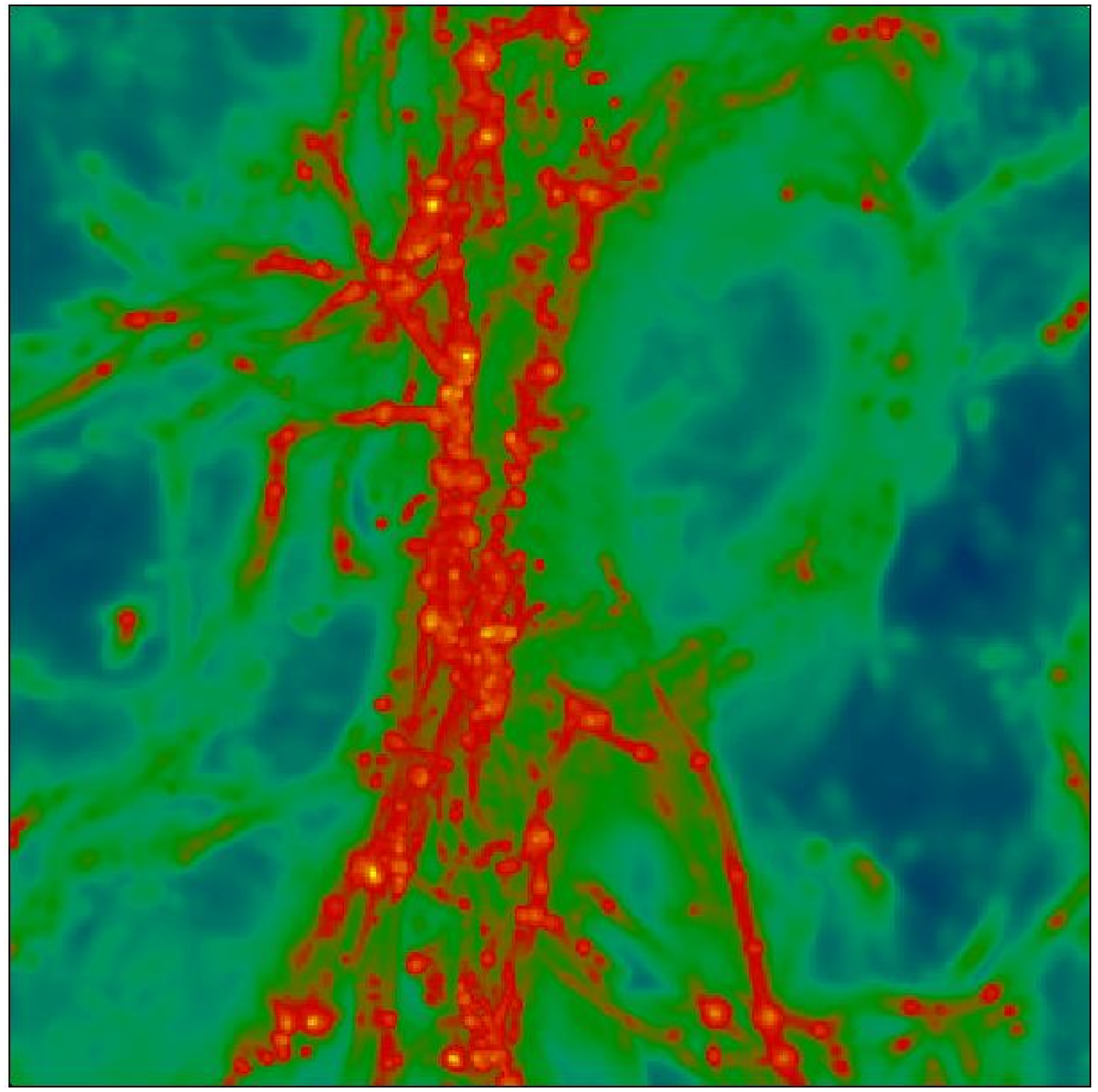}}
\put(5.7,7.9){\includegraphics[width=5.6cm,height=5.6cm]{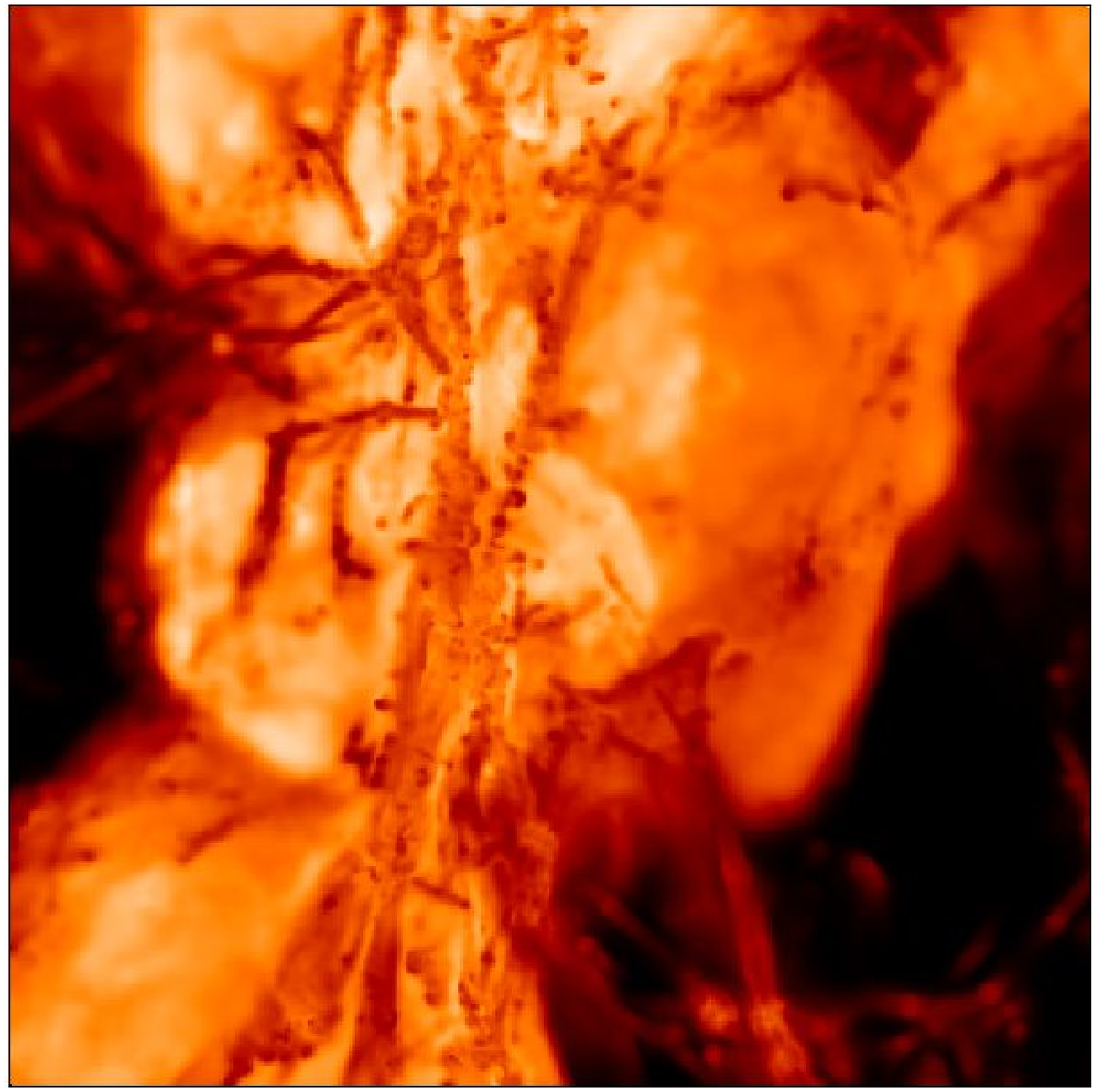}}
\put(11.4,7.9){\includegraphics[width=5.6cm,height=5.6cm]{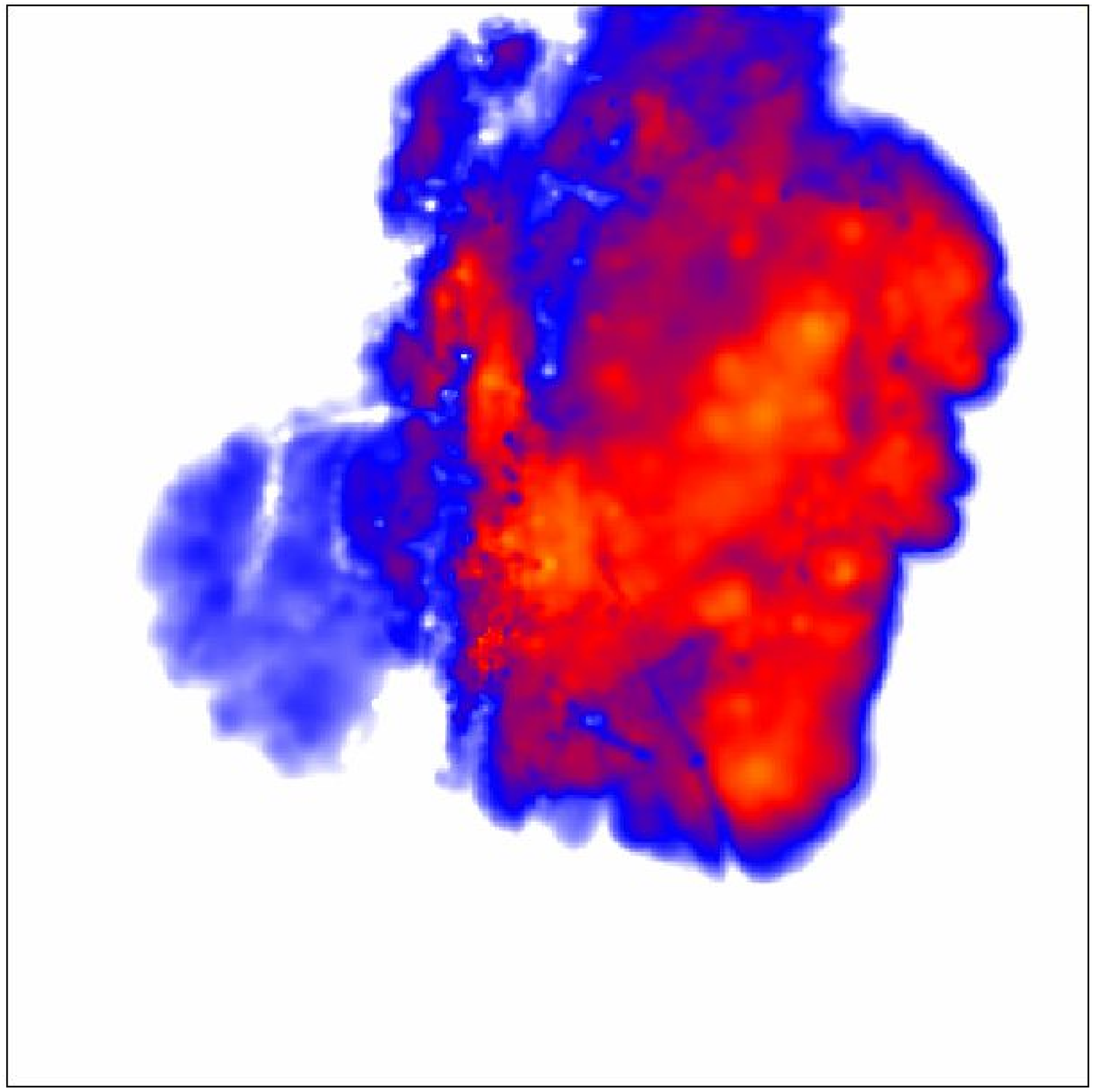}}
\put(0.0,2.1){\includegraphics[width=5.6cm,height=5.6cm]{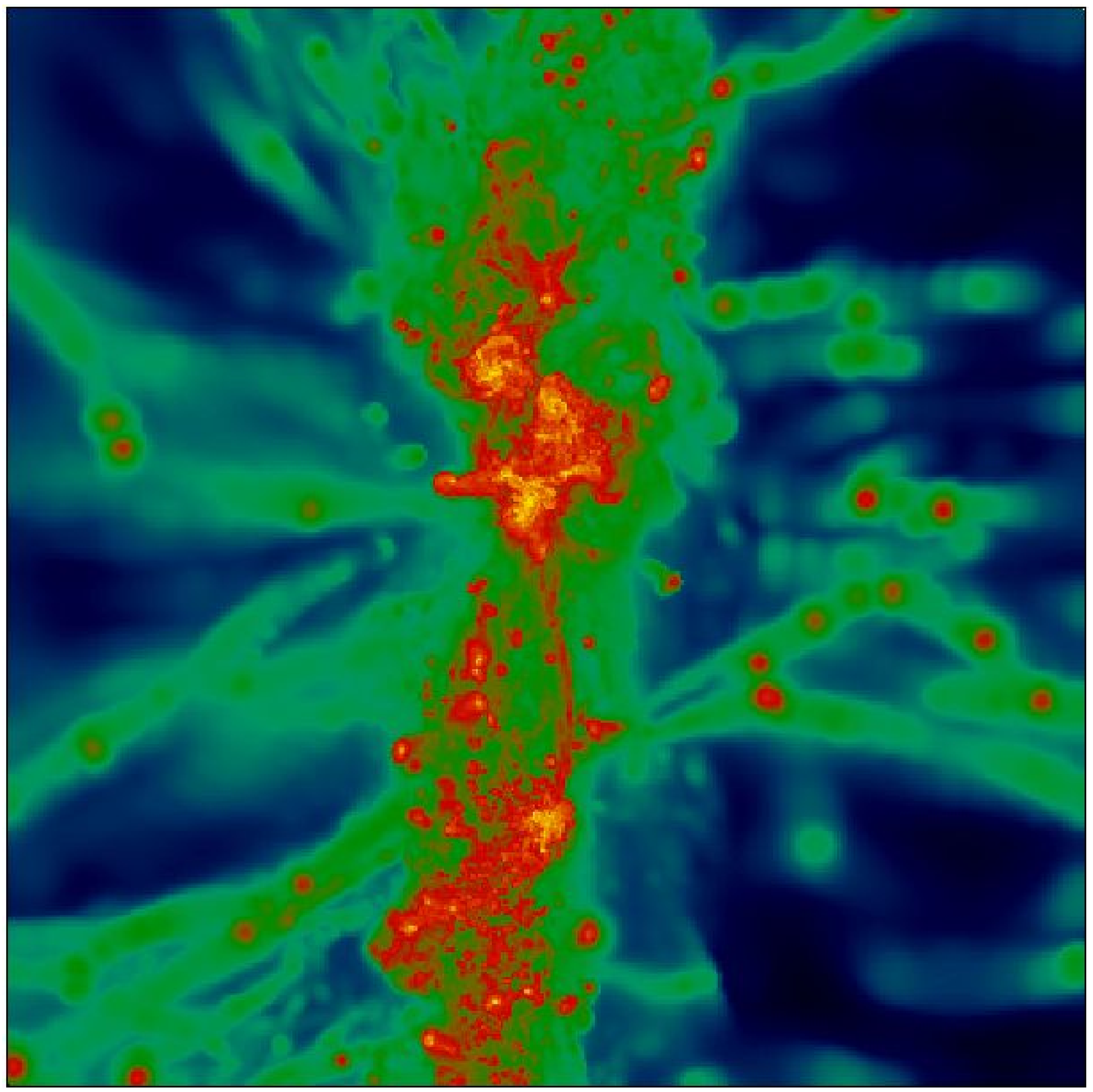}}
\put(5.7,2.1){\includegraphics[width=5.6cm,height=5.6cm]{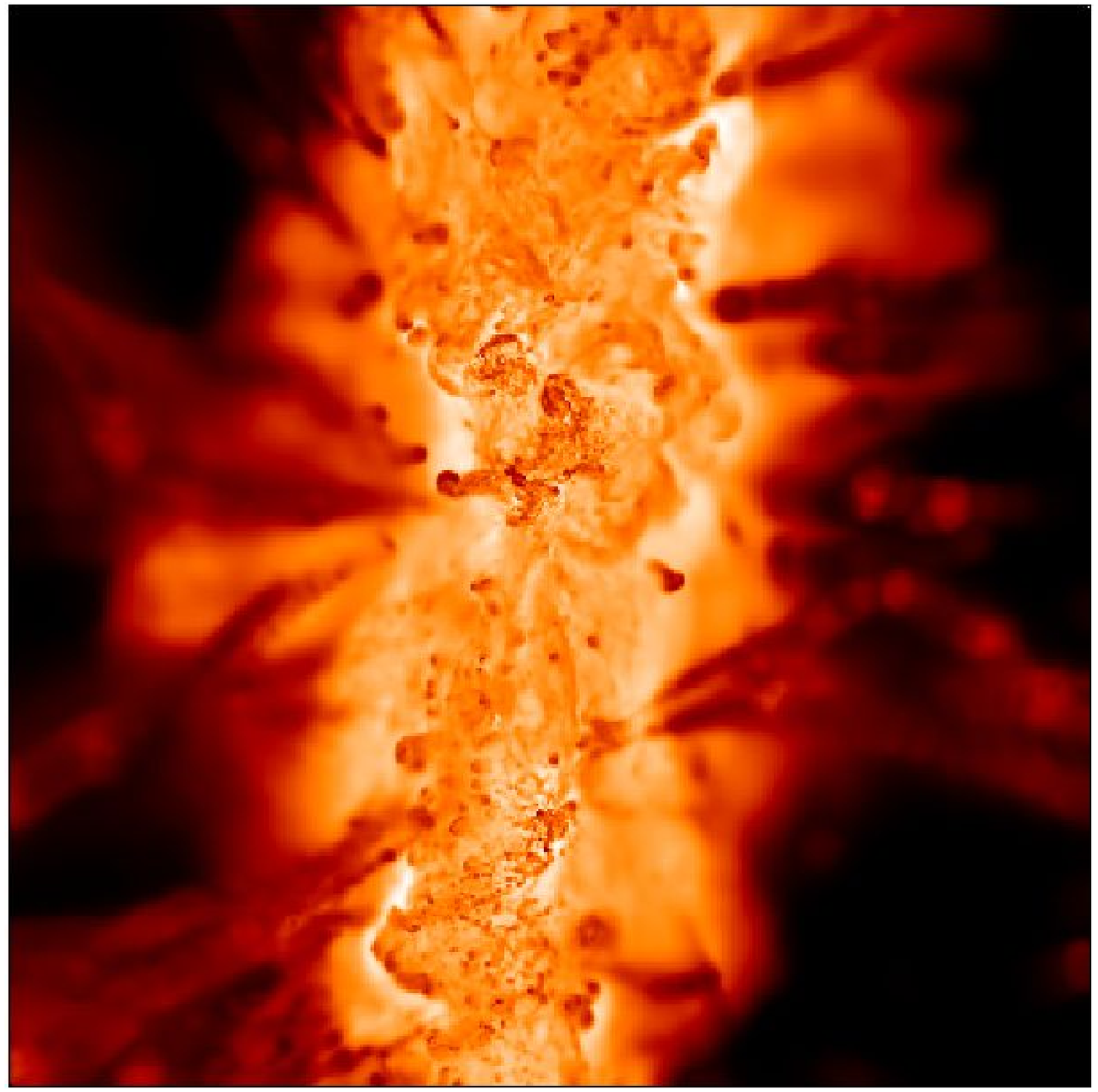}}
\put(11.4,2.1){\includegraphics[width=5.6cm,height=5.6cm]{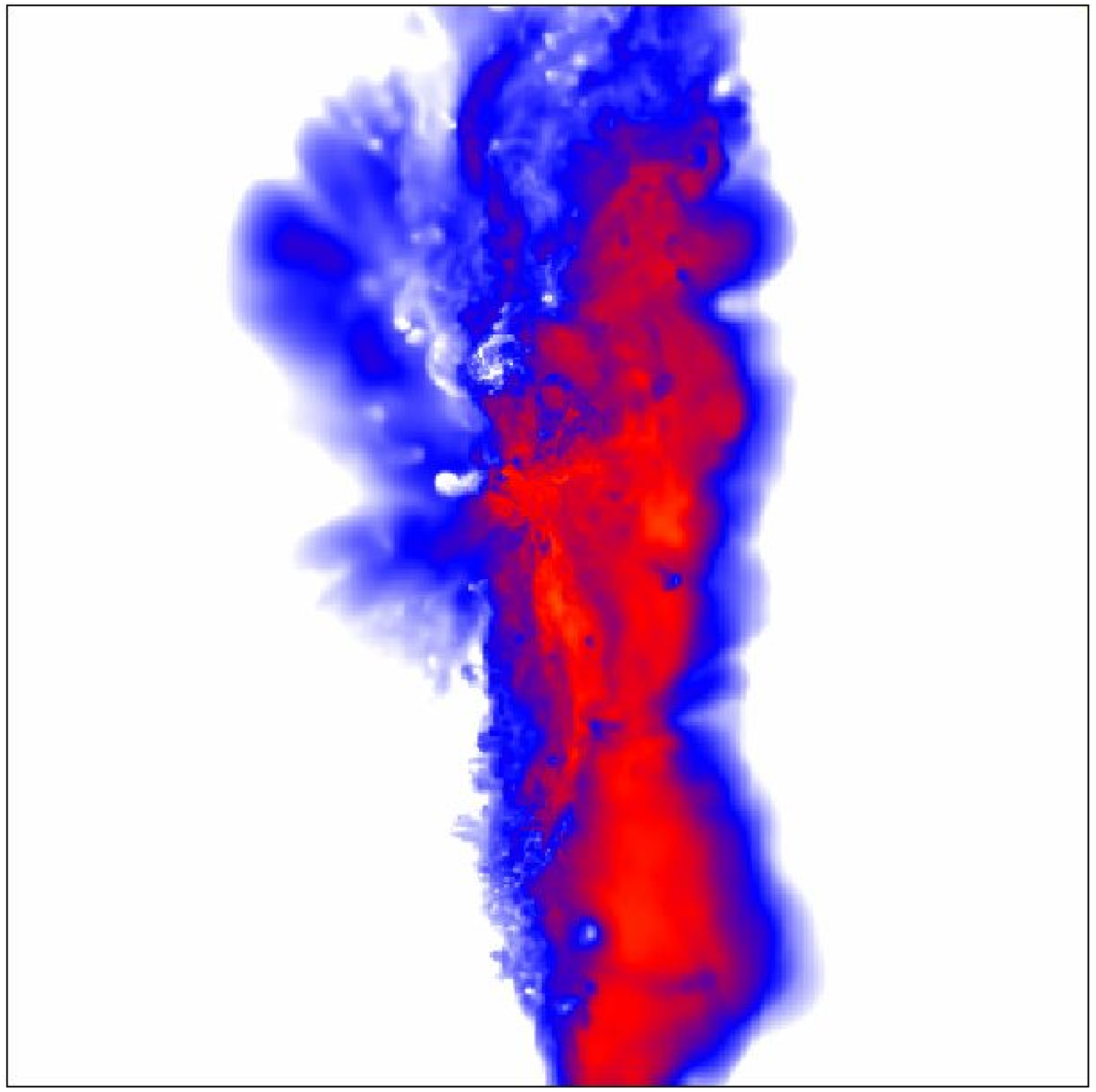}}
\put(3.1,2.1){\includegraphics[width=2.5cm,height=2.5cm]{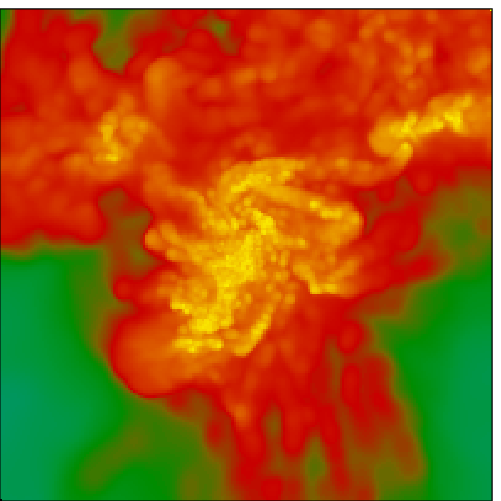}}
\put(8.8,2.1){\includegraphics[width=2.5cm,height=2.5cm]{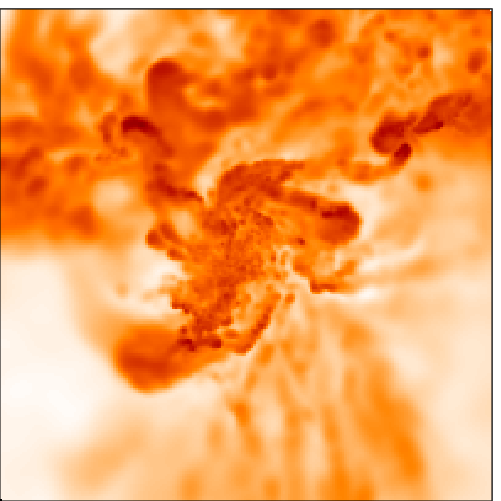}}
\put(14.5,2.1){\includegraphics[width=2.5cm,height=2.5cm]{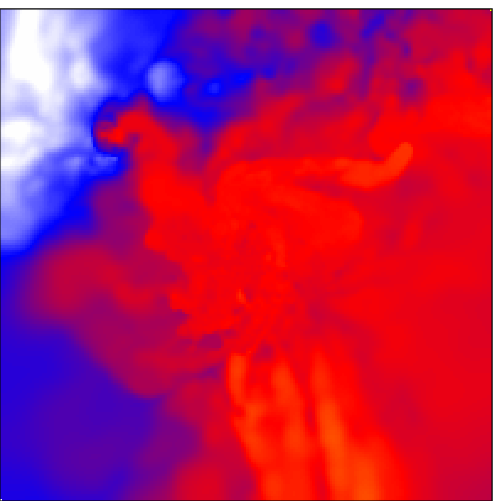}}
\put(0.0,0.0){\includegraphics[width=5.6cm,height=2.0cm]{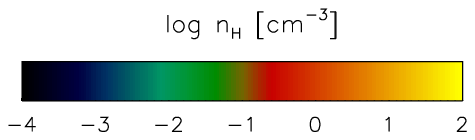}}
\put(5.7,0.0){\includegraphics[width=5.6cm,height=2.0cm]{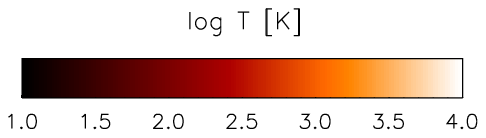}}
\put(11.4,0.0){\includegraphics[width=5.6cm,height=2.0cm]{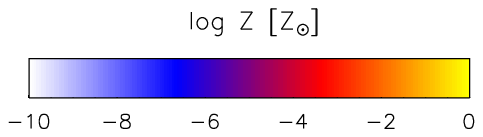}}
\end{picture}}
\caption{From left to right: the density-squared weighted average of the hydrogen density, temperature, and metallicity along the line of sight in the central $100~{\rm kpc}$ (comoving) of Sim~A. From top to bottom: a time series showing the simulation $15$, $100$, and $300~{\rm Myr}$ after the SN explosion. The inlays show the central $10~{\rm kpc}$ (comoving) of the nascent galaxy. The metals are initially distributed by the bulk motion of the SN remnant, and later by turbulent motions induced by photoheating from other stars and the virialization of the galaxy. As can be seen in the inlays, the gas within the newly formed galaxy is highly enriched.}
\end{center}
\end{figure*}

\begin{figure*}
\begin{center}
\resizebox{17cm}{20.9cm}
{\unitlength1cm
\begin{picture}(17,20.9)
\put(0.0,19.4){\includegraphics[width=17.0cm,height=1.5cm]{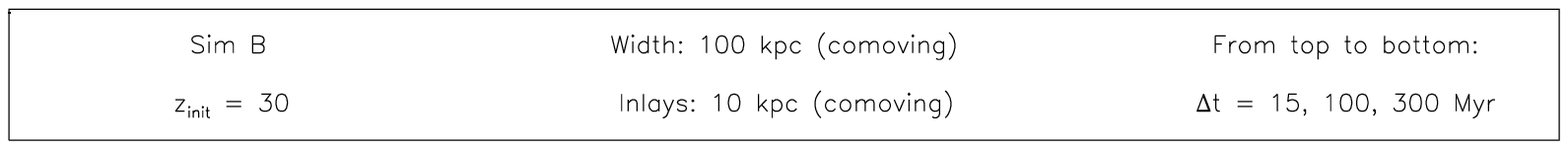}}
\put(0.0,13.7){\includegraphics[width=5.6cm,height=5.6cm]{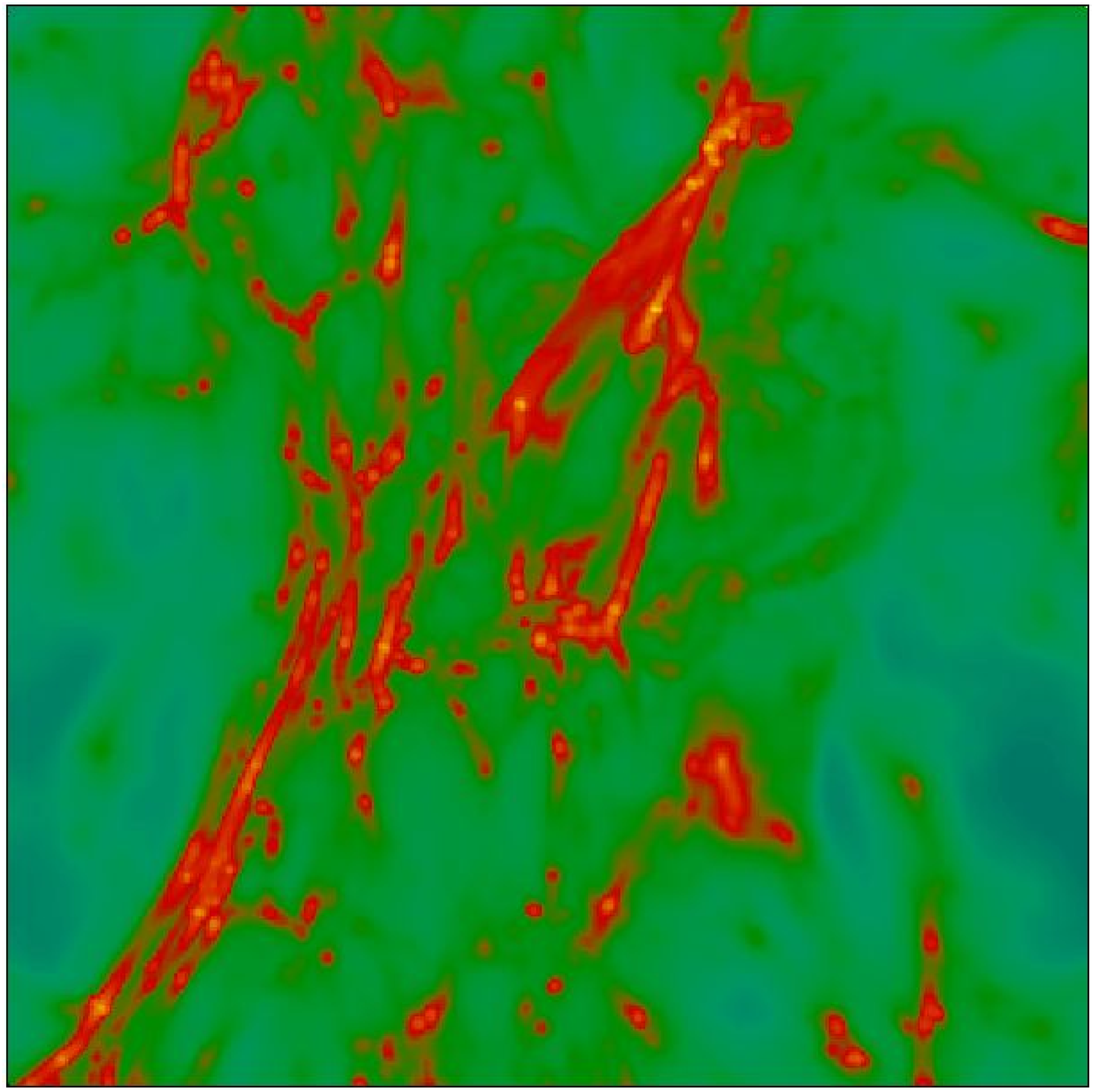}}
\put(5.7,13.7){\includegraphics[width=5.6cm,height=5.6cm]{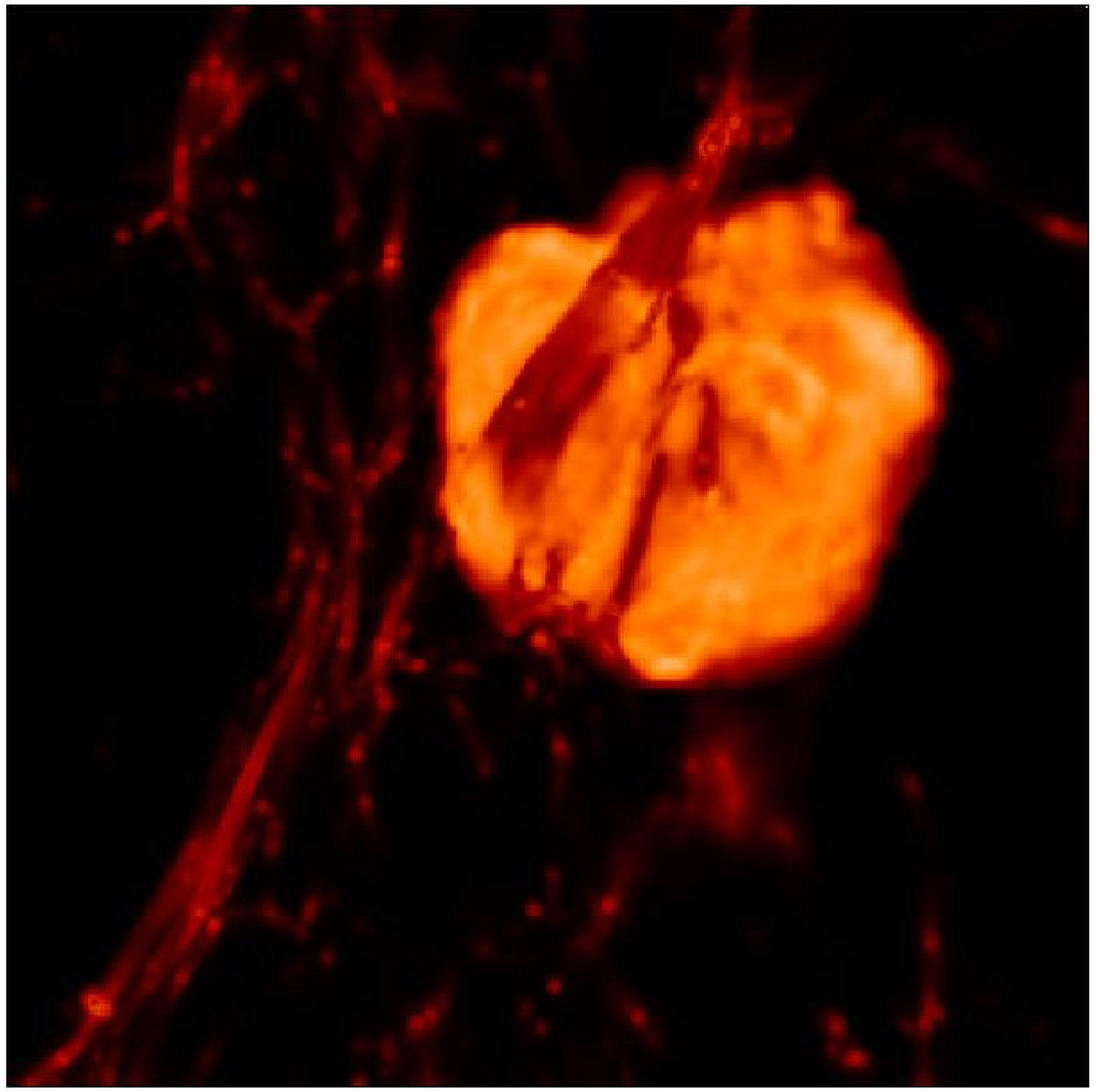}}
\put(11.4,13.7){\includegraphics[width=5.6cm,height=5.6cm]{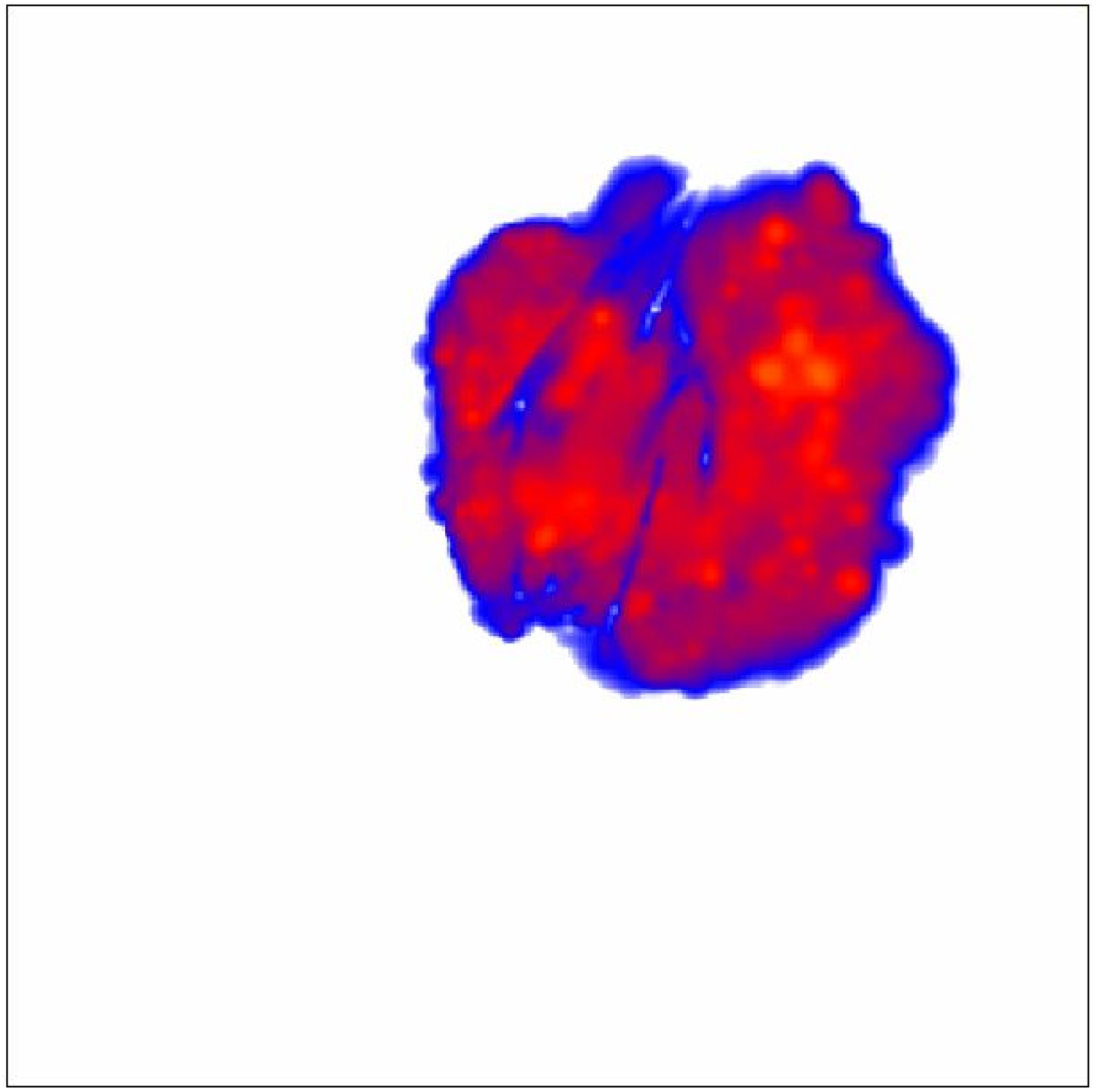}}
\put(0.0,7.9){\includegraphics[width=5.6cm,height=5.6cm]{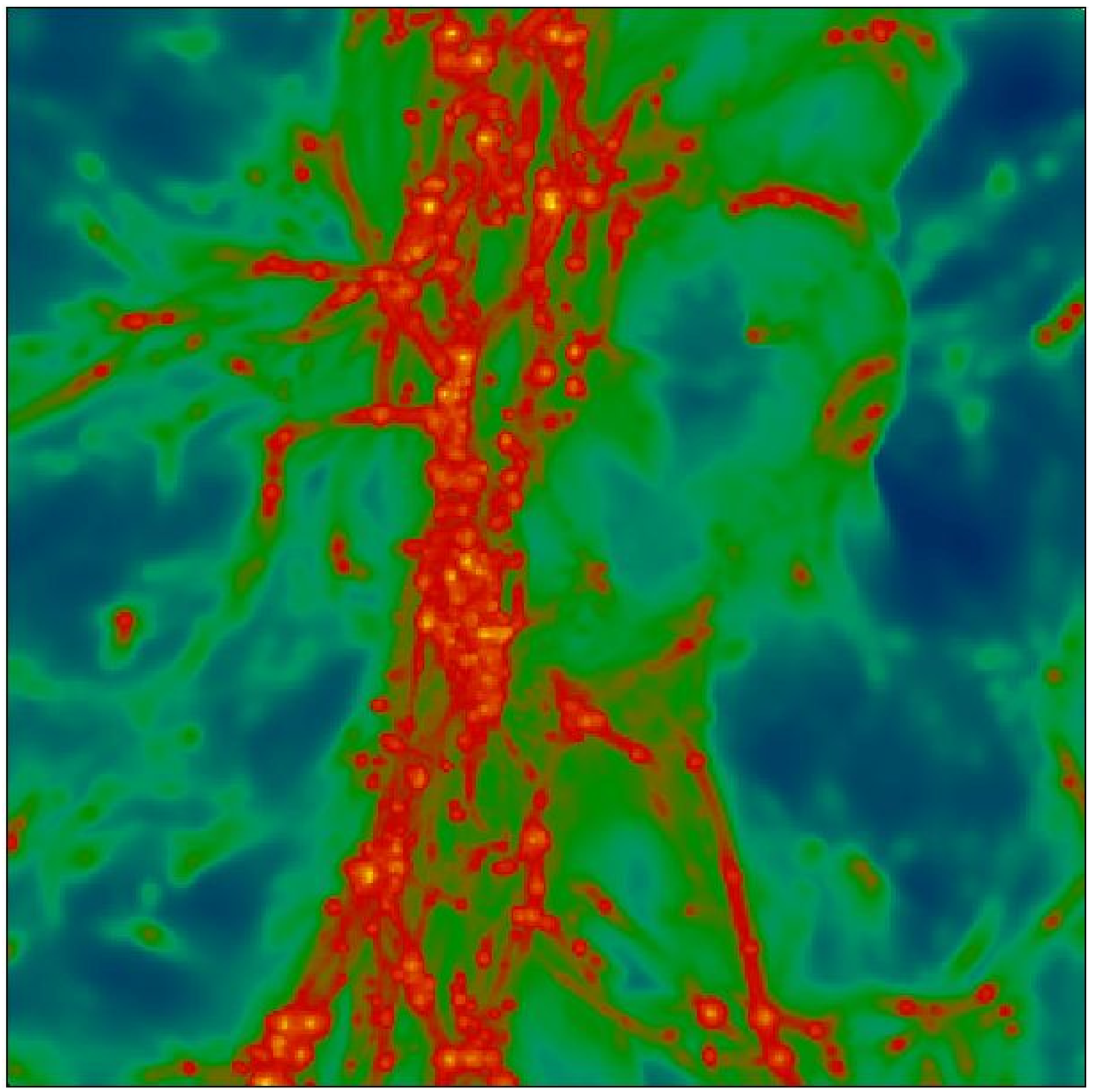}}
\put(5.7,7.9){\includegraphics[width=5.6cm,height=5.6cm]{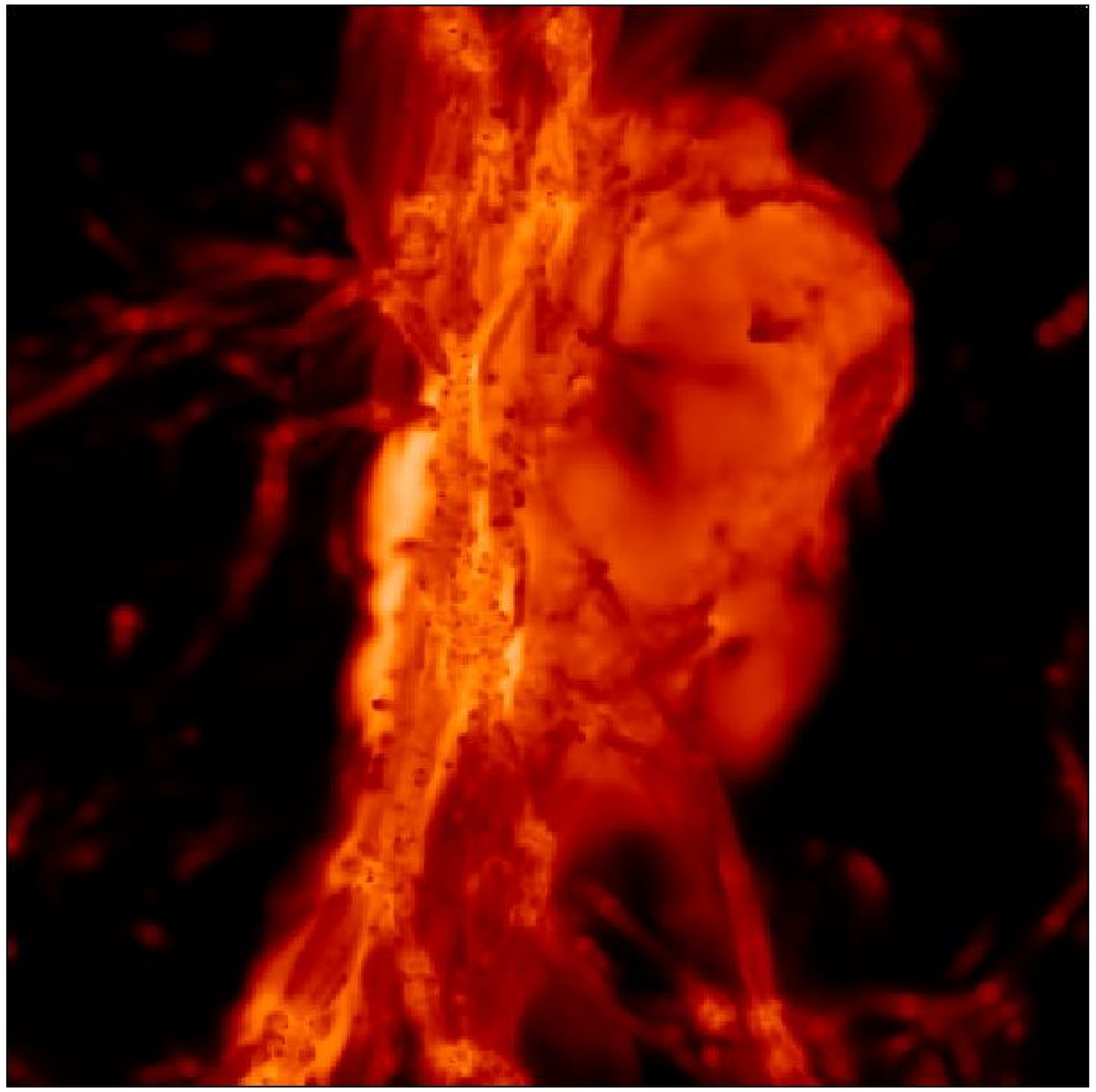}}
\put(11.4,7.9){\includegraphics[width=5.6cm,height=5.6cm]{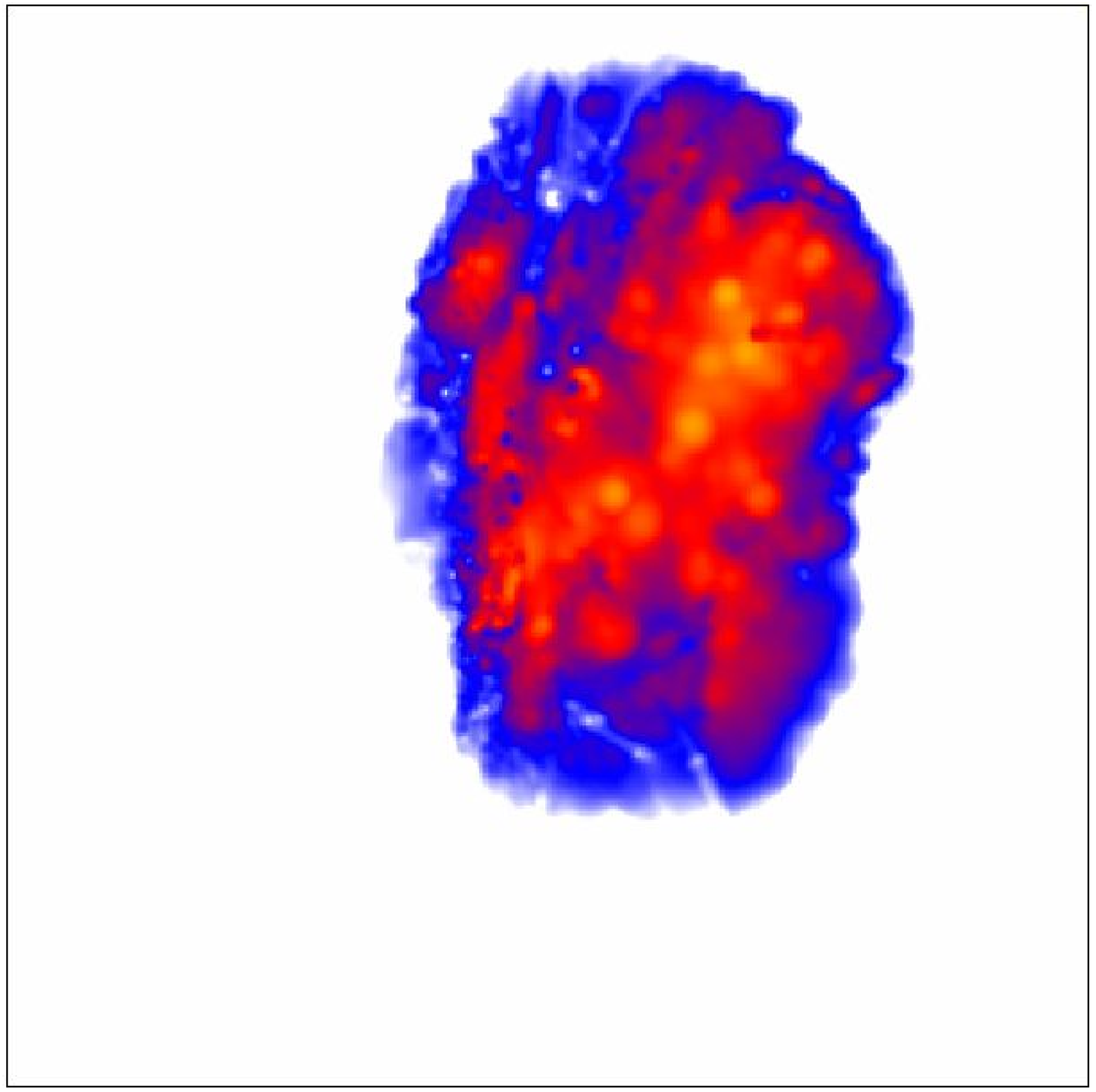}}
\put(0.0,2.1){\includegraphics[width=5.6cm,height=5.6cm]{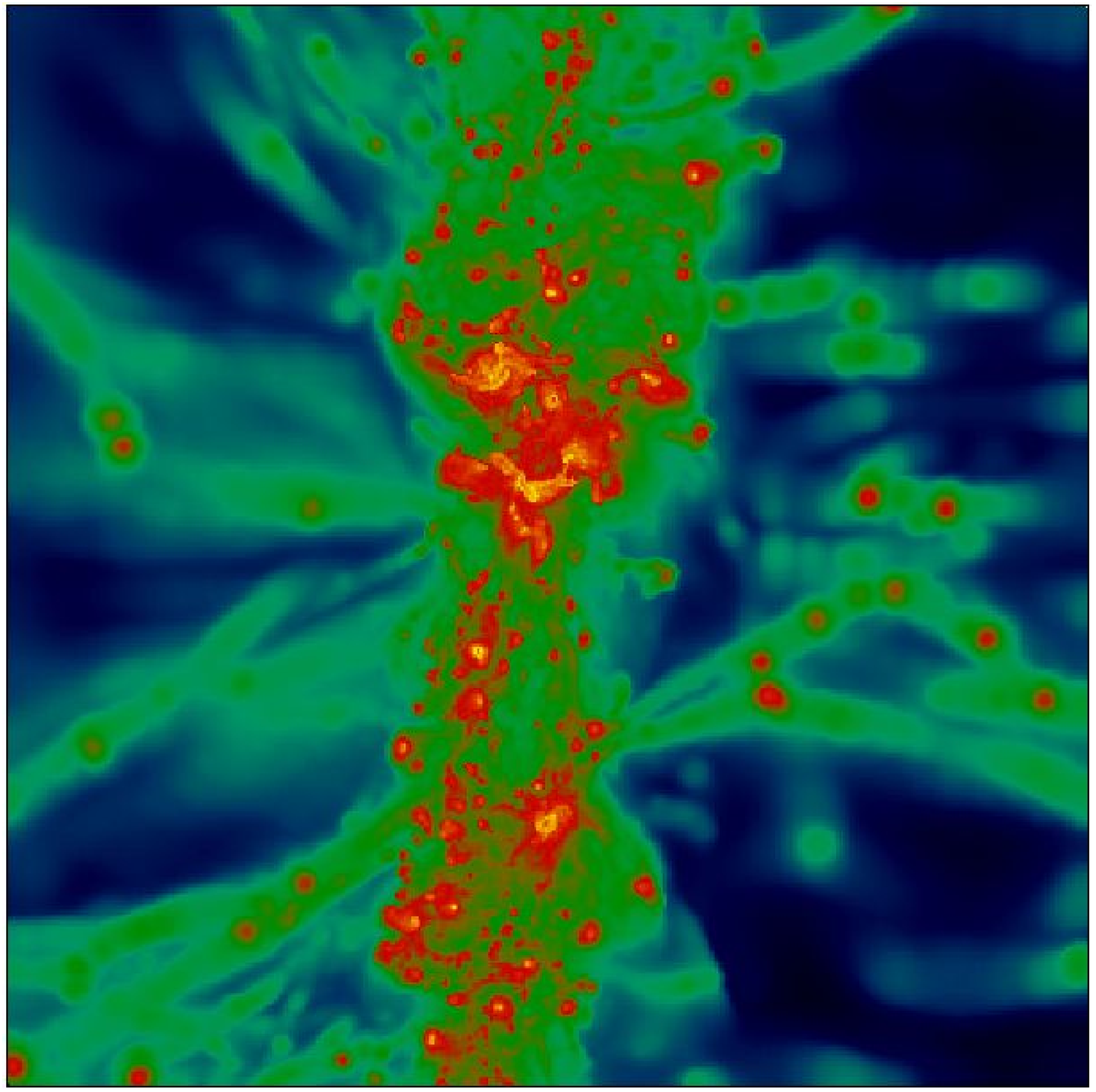}}
\put(5.7,2.1){\includegraphics[width=5.6cm,height=5.6cm]{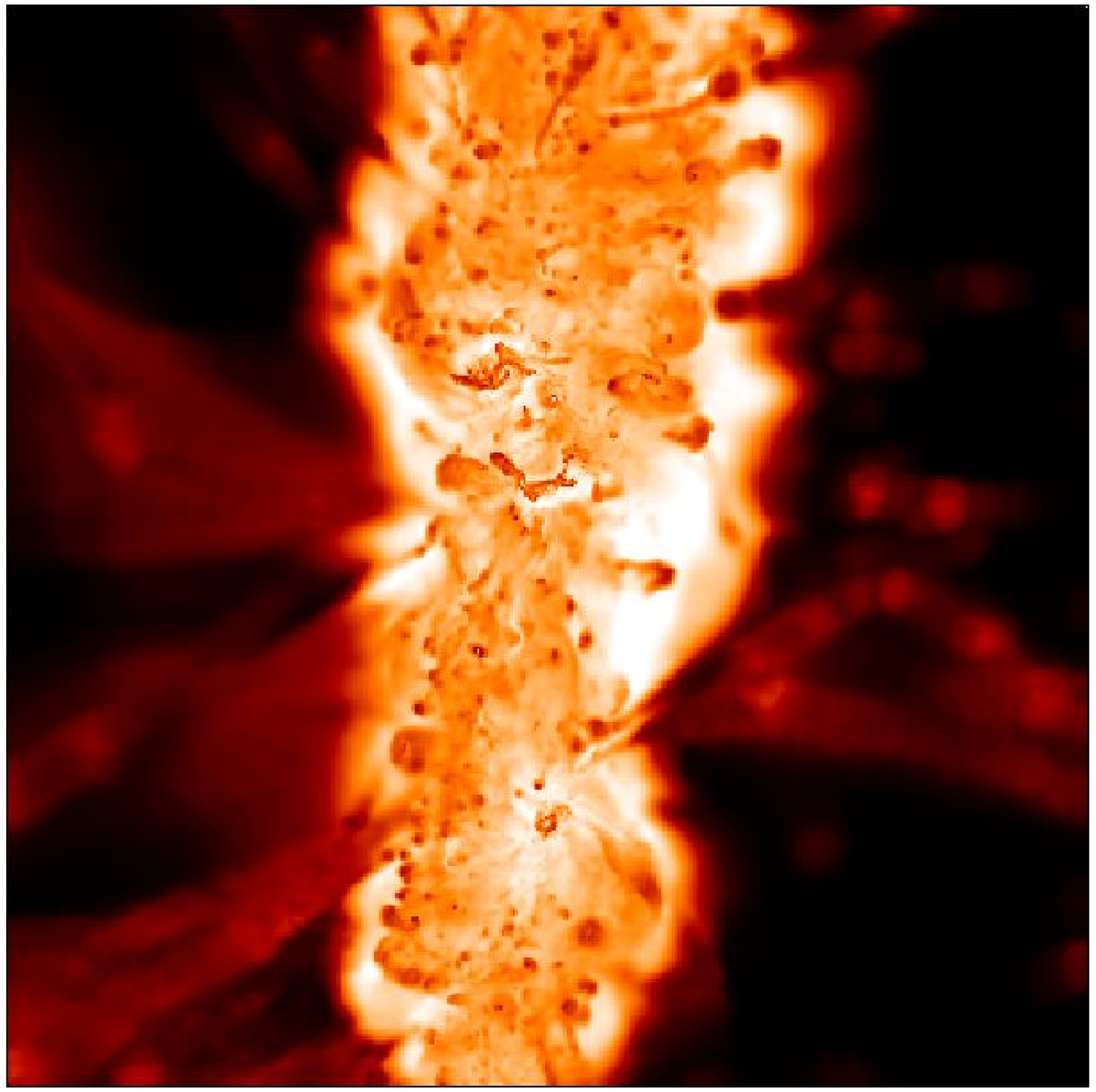}}
\put(11.4,2.1){\includegraphics[width=5.6cm,height=5.6cm]{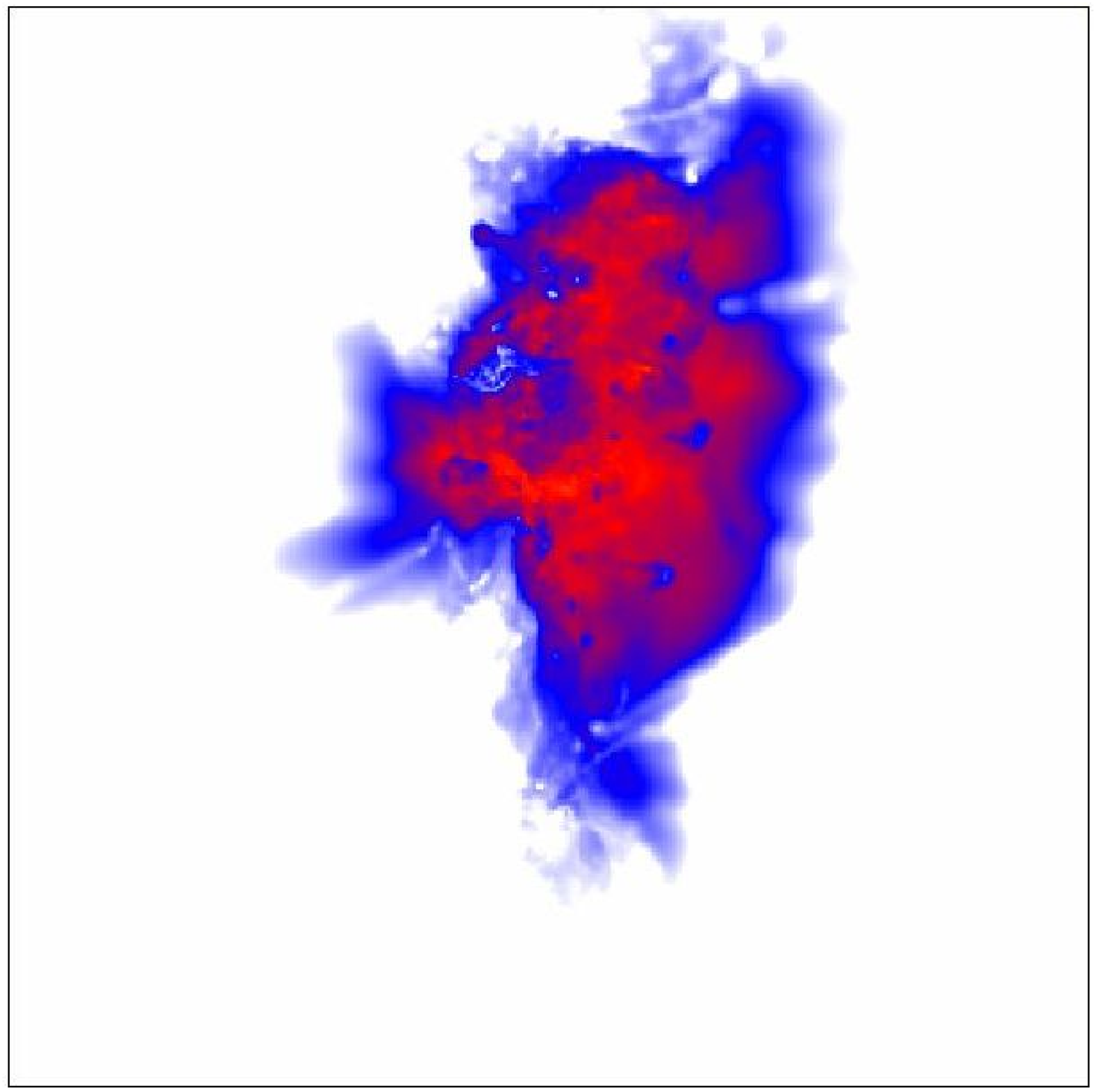}}
\put(3.1,2.1){\includegraphics[width=2.5cm,height=2.5cm]{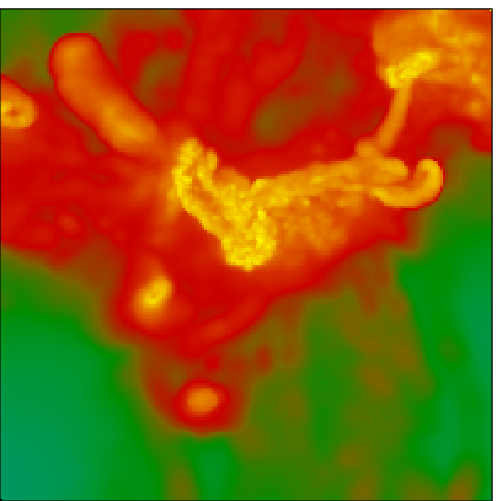}}
\put(8.8,2.1){\includegraphics[width=2.5cm,height=2.5cm]{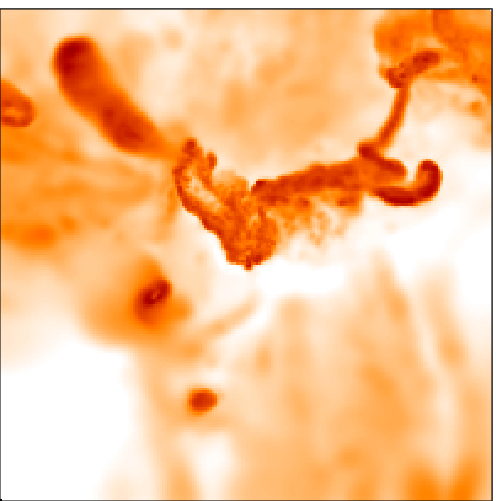}}
\put(14.5,2.1){\includegraphics[width=2.5cm,height=2.5cm]{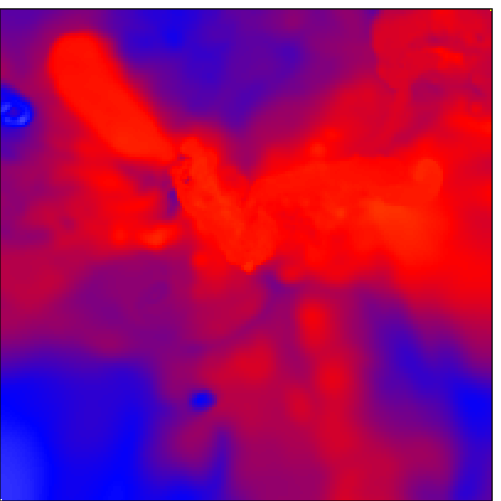}}
\put(0.0,0.0){\includegraphics[width=5.6cm,height=2.0cm]{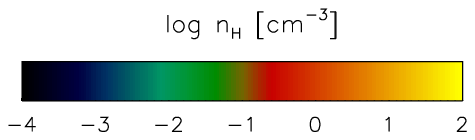}}
\put(5.7,0.0){\includegraphics[width=5.6cm,height=2.0cm]{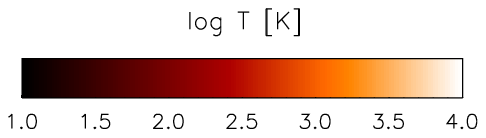}}
\put(11.4,0.0){\includegraphics[width=5.6cm,height=2.0cm]{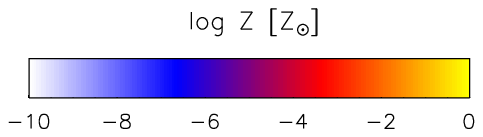}}
\end{picture}}
\caption{Same as in Figure~1, but without additional photoheating from nearby stars (Sim~B). This results in a significantly smaller enriched region as well as less efficient mixing. The gas within the galaxy is more concentrated, and we find that virialization shocks are hotter and more pronounced than in Sim~A, where the gas shocks at larger distances from the halo, so that the infall velocity is smaller.}
\end{center}
\end{figure*}

\begin{figure}
\begin{center}
\resizebox{8cm}{13.2cm}
{\unitlength1cm
\begin{picture}(8,13.2)
\put(0.0,12.4){\includegraphics[width=8cm,height=0.8cm]{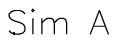}}
\put(0.0,7.6){\includegraphics[width=8cm,height=4.8cm]{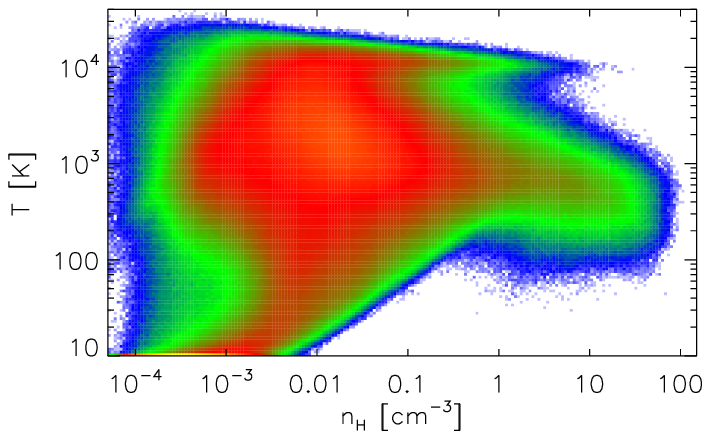}}
\put(0.0,6.8){\includegraphics[width=8cm,height=0.8cm]{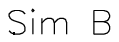}}
\put(0.0,2.0){\includegraphics[width=8cm,height=4.8cm]{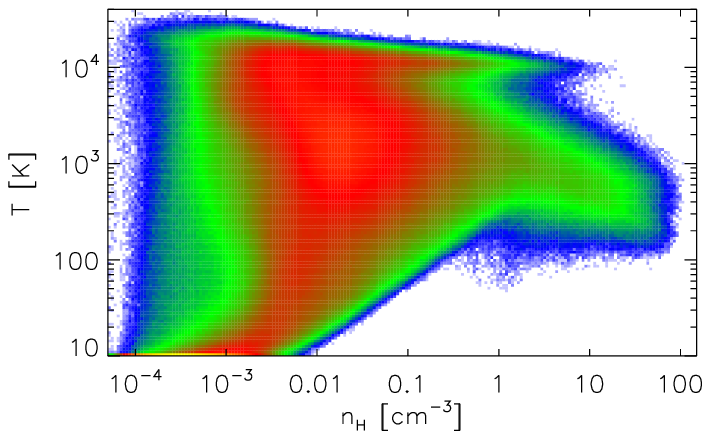}}
\put(0.0,0.0){\includegraphics[width=8cm,height=2.0cm]{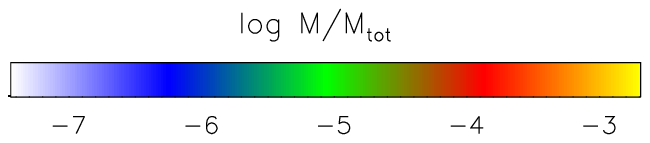}}
\end{picture}}
\caption{Mass-weighted distribution of gas in density and temperature space within the high resolution region at $z\simeq 10$, color-coded according to its mass fraction. Most of the gas is shock-heated to the virial temperature and cools once the density exceeds $\sim 1~{\rm cm}^{-3}$, but a significant fraction of the gas in Sim~A is also relic H~{\sc ii} region gas that has cooled to $T\simeq 1000~{\rm K}$ and resides in the low-density IGM at $n_{\rm H}\simeq 10^{-3}-10^{-4}~{\rm cm}^{-3}$.}
\end{center}
\end{figure}

Concomitant to the onset of star formation in other minihalos, the potential well of the nascent galaxy assembles at the center of a massive filament (see bottom panels of Figures~1 and 2). Accreted halos are subject to tidal stripping and rapidly lose their mass, similar to the disruption of satellite galaxies entering the halo of the Milky Way. In some cases, they complete a few orbits before being disrupted, or undergo a major merger. Interestingly, virial shocks around halos in Sim~B are hotter and more distinct than in Sim~A. The additional photoheating in Sim~A increases the pressure at larger radii, reducing the infall velocity and, in consequence, the post-shock temperature of the gas. For the same reason, structures in Sim~B are generally more pronounced, which is particularly true at the center of the galaxy (see inlays).

\subsection{Distribution of metals}

During the first few million years after the SN explosion, the metals are distributed into the IGM by the bulk motion of the blast wave, which leads to a tight correlation between temperature and metallicity. This becomes evident from the top panels shown in Figures~1 and 2. Once the SN remnant stalls, mixing is facilitated by agents that act on various scales. On the largest scales, filamentary accretion dominates \citep[e.g.,][]{wa07b,greif08}, while on smaller scales photoheating by neighboring stars as well as dynamical interactions associated with the virialization of the galaxy become important. As can be seen in the middle and bottom panels of Figures~1 and 2, these all interact to create a complex hierarchy of turbulent motions that efficiently mix the gas. In Figure~4, we show the resulting density-metallicity relation for the three output times in Figures~1 and 2. At early times, a distinct correlation arises: Hot, underdense regions of the IGM tend to be enriched to $Z\sim 10^{-3}~Z_{\odot}$ or higher, while the densest regions, such as nearby minihalos, remain almost devoid of metals. Once the potential well of the galaxy assembles, metal-rich gas accumulates at high densities and leads to a flattening of the relation.

A more detailed view of the metal distribution is shown in Figures~5 and 6, where we plot the enriched mass as a function of gas overdensity and metallicity. Early on, the initial ejecta have not propagated very far and the metals occupy a small region in density and metallicity space, located at the relic H~{\sc ii} region density of $\sim 0.1~{\rm cm}^{-3}$ and at supersolar metallicities. Over time, the SN remnant then distributes its metals to the surrounding medium and the average metallicity decreases, while the total enriched mass increases. Once the potential well of the galaxy has become deep enough, a large fraction of the hot, metal-rich gas residing in the IGM has recollapsed to high densities. By $z\simeq 10$, the enriched mass contained within the densest parcels of gas in the galaxy has increased to roughly $10^{5}~M_{\odot}$, with a clear peak in the distribution at $Z\sim 10^{-3}~Z_{\odot}$. Interestingly, this exceeds any critical metallicity quoted in the literature \citep{bl03a,schneider06}, and we expect that a stellar cluster with a normal IMF will form \citep[e.g.,][]{omukai05,schneider06,cgk08,clark09}. It therefore appears that a single PISN is sufficient to trigger a transition from Pop~III to Pop~II star formation.

Finally, we note that the additional photoheating in Sim~A has an interesting effect on the distribution of metals. As is evident from Figures~1 and 2, the enriched region is generally larger and better mixed in Sim~A. In some cases, the photoheating even ejects significant amounts of enriched gas out of the potential well of the galaxy, which can be seen in the case of the metal bubble extending to the lower left of the simulation box. This behavior is intriguing in light of claims that the IGM is substantially enriched at high redshifts \citep{sc96,schaye03}, although this is generally attributed to SN feedback instead of photoheating, which acts on objects with small circular velocities \citep{tw96,dijkstra04}. Comparing Figures~5 and 6, we also find more quantitative evidence of this effect. From top left to bottom right, an increasing amount of enriched gas accumulates at low densities, which then accretes onto the nascent galaxy.

\subsection{Metal Cooling}

In both simulations, we have included fine-structure cooling for C, O and Si at low temperatures and collisional excitation and recombination cooling at high temperatures. Depending on metallicity, the latter can dominate over H and He cooling at very early times. Such a phase exists during the first few million years after the SN explosion, when the temperature of the SN remnant exceeds $10^{5}~{\rm K}$ and the metallicity within the remnant is well above solar. By performing test runs with and without metal line cooling in a smaller simulation box, we have found that metals temporarily enhance the net cooling rate by a factor of a few. However, even for the case of a PISN, the influence on the dynamical evolution of the SN remnant remains small.

\begin{figure}
\begin{center}
\includegraphics[width=8cm]{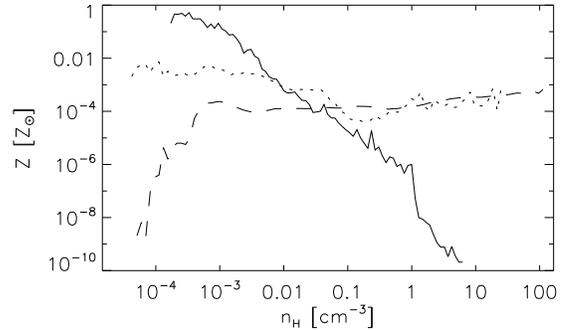}
\caption{Mass-weighted, average metallicity as a function of gas density within the high resolution region, shown $\simeq 15$ (solid line), $100$ (dotted line), and $300~{\rm Myr}$ (dashed line) after the SN explosion (Sim~A). At early times, a distinct correlation is evident: underdense region are highly enriched, while overdense region remain largely pristine. Once the potential well of the galaxy assembles, metal-rich gas becomes dense and the relation flattens (see also Figures~5 and 6).}
\end{center}
\end{figure}

\begin{figure*}
\begin{center}
\resizebox{17cm}{18.2cm}
{\unitlength1cm
\begin{picture}(17,18.2)
\put(0.0,17.0){\includegraphics[width=17.0cm,height=1.2cm]{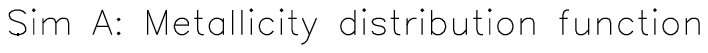}}
\put(0.0,0.0){\includegraphics[width=17.0cm,height=17.0cm]{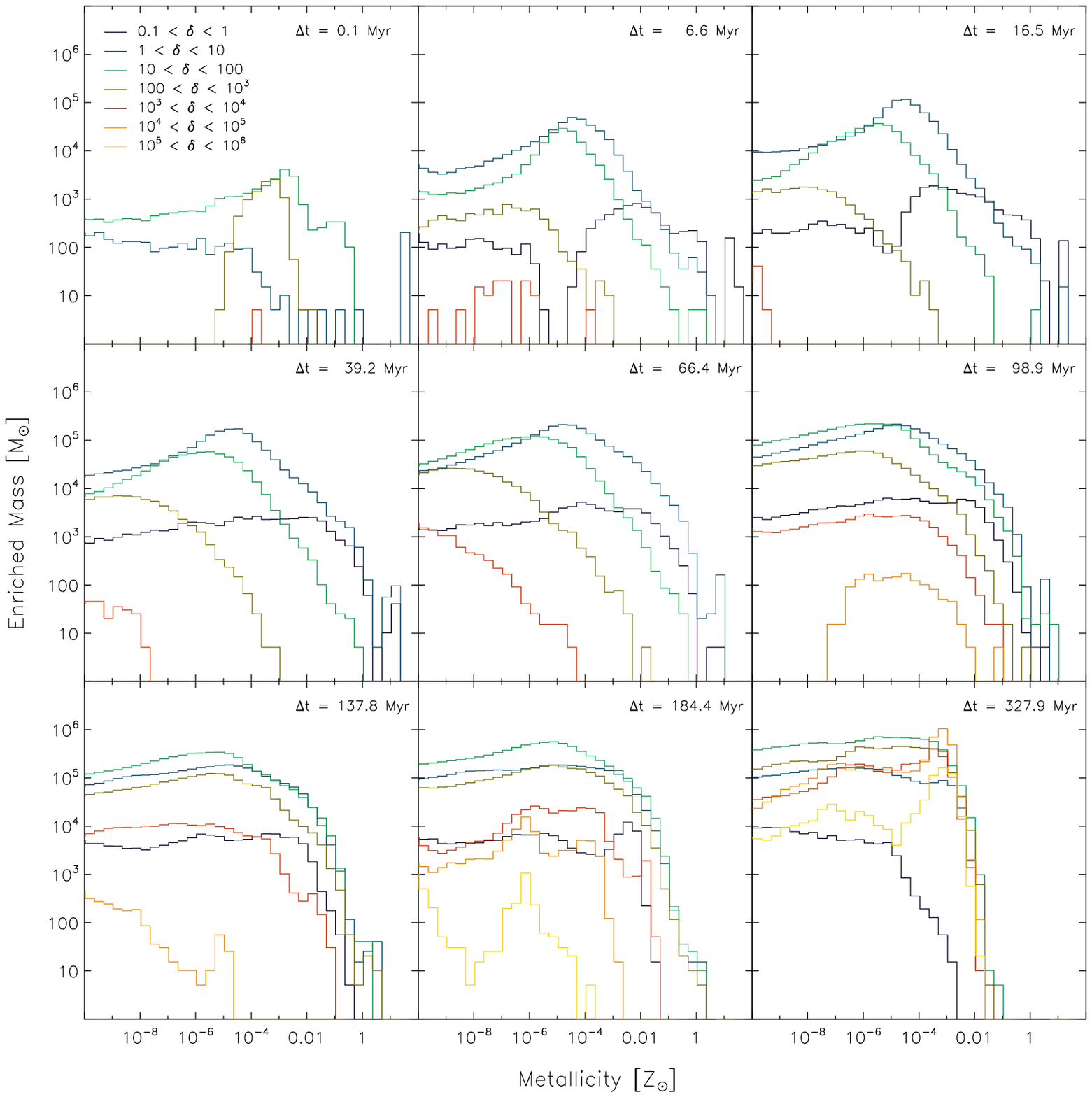}}
\end{picture}}
\caption{Enriched mass as a function of metallicity and gas overdensity, shown for different times after the SN explosion (Sim~A). The overdensity is color-coded according to the legend shown in the top left panel. At early times, the metals remain confined to the initial stellar ejecta, located at a density of $\sim 0.1~{\rm cm}^{-3}$ and at supersolar metallicity. Later on, the gas mixes with the surrounding medium and the total enriched mass increases, while the average metallicity decreases. Once the potential well of the galaxy assembles, the gas recollapses to high densities and becomes available for star formation. At the final output time, the metallicity distribution within the galaxy peaks at $Z\sim 10^{-3}~Z_{\odot}$.}
\end{center}
\end{figure*}

\begin{figure*}
\begin{center}
\resizebox{17cm}{18.2cm}
{\unitlength1cm
\begin{picture}(17,18.2)
\put(0.0,17.0){\includegraphics[width=17.0cm,height=1.2cm]{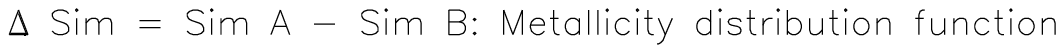}}
\put(0.0,0.0){\includegraphics[width=17.0cm,height=17.0cm]{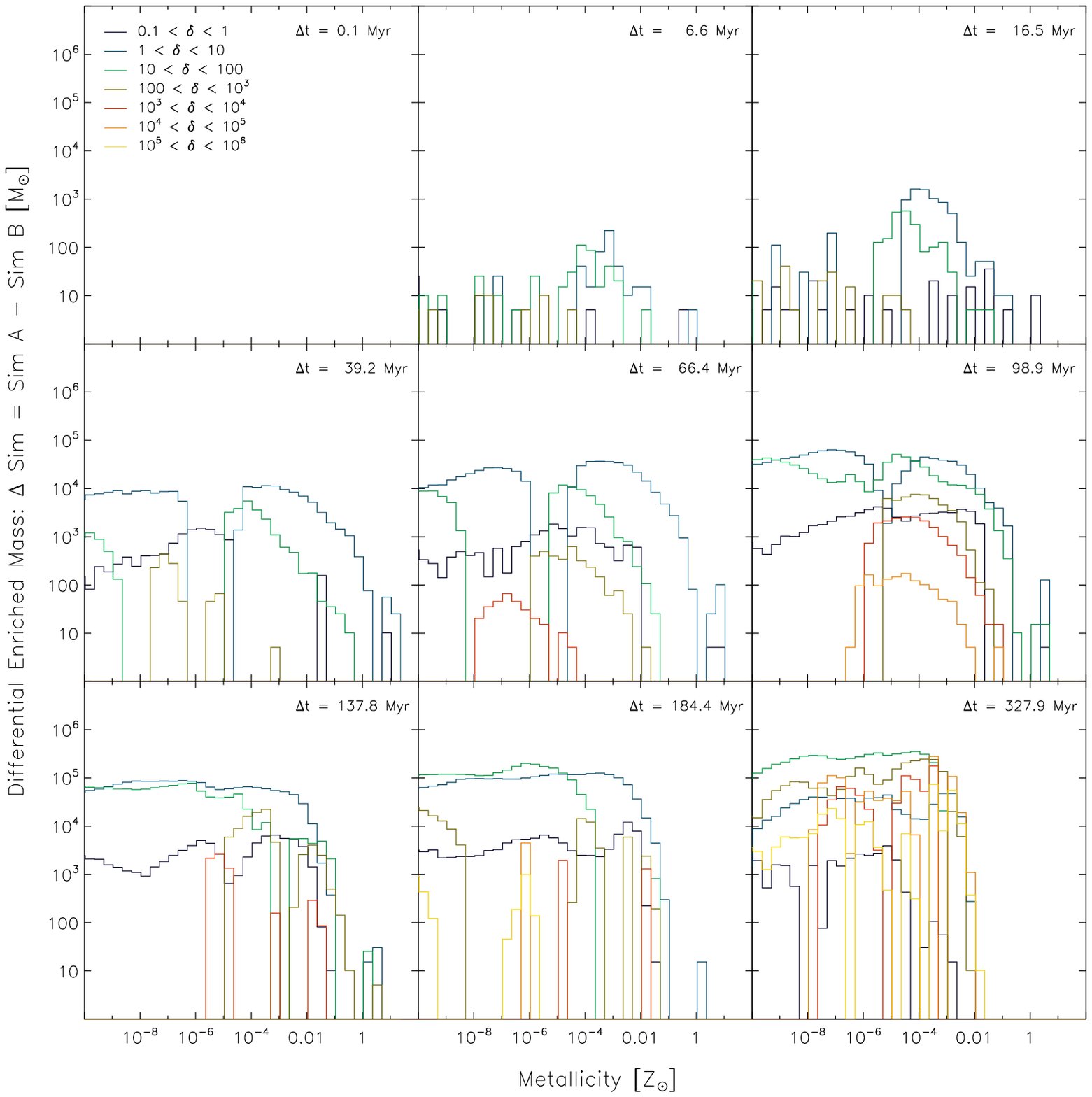}}
\end{picture}}
\caption{Same as in Figure~5, except that we show the differential enriched mass obtained by subtracting the metal content in each metallicity bin in Sim~B from that in Sim~A. From top left to bottom right, the amount of enriched gas ejected into the IGM by the additional photoheating in Sim~A increases. At the final output time, a large fraction of this gas has accreted onto the galaxy.}
\end{center}
\end{figure*}

At temperatures below $10^{4}~{\rm K}$, fine-structure cooling provided by heavy elements takes over. However, we find that once the gas has cooled to the regime where fine-structure cooling becomes important, the metallicity has dropped such that cooling is instead dominated by H$_{2}$. This was already found by \citet{jappsen07,jappsen09a}, who argued that dust is responsible for changing the cooling and fragmentation properties of the gas. Since cooling via dust grains kicks in at densities much higher than we can resolve here, a detailed treatment is beyond the scope of this work, although we expect that it will play a crucial role for the further evolution of the gas \citep[e.g.,][]{omukai05}. We note that the relative importance of the different cooling channels could be altered in the presence of a strong Lyman--Werner (LW) radiation background, which may lead to the dissociation of H$_2$ \citep[e.g.,][]{hrl97,jgb07,ahn09}.

\section{Second-generation star formation}

To elucidate the pathway to the first galaxies, an essential ingredient is to characterize the conditions in the gas clouds out of which the second generation of stars form \citep[e.g.,][]{bromm09}. We in particular need to quantify the metal content in such clouds. This applies to the first galaxy itself, but also to neighboring minihalos which collapse during the assembly process. It is generally believed that minihalos can only host Pop~III stars, as self-enrichment is unlikely, and transport of metals out of the IGM into pre-condensed halos might be severely limited \citep{cr08}. Here, we can address these questions within their proper cosmological setting. We note that we use the terminology to classify early stellar populations, in particular the distinction between Pop~III.1 and III.2, recently introduced by \citet{mt08}. The former refers to metal-free stars whose formation is entirely unaffected by any stellar feedback, and is instead governed by cosmological initial conditions, whereas the latter denotes metal-free stars that form under the influence of feedback effects.

\begin{figure}
\begin{center}
\includegraphics[width=8cm]{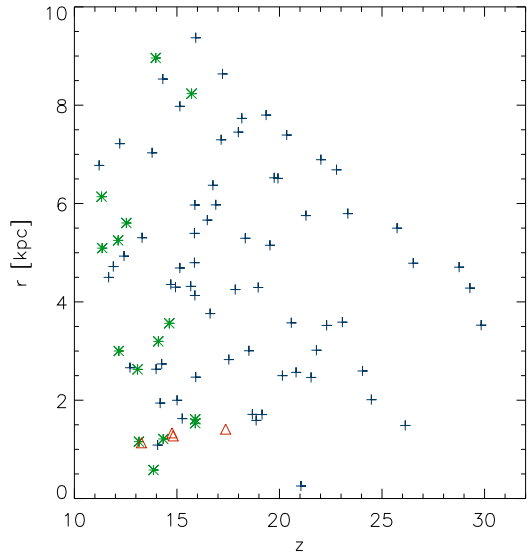}
\caption{All star-forming minihalos in the high resolution region as a function of redshift and distance from the nascent galaxy (Sim~A). Crosses denote undisturbed halos that will form Pop~III.1 stars, star symbols denote halos that have been photoionized and will likely form Pop~III.2 stars, and triangles denote halos that have been enriched to $Z>10^{-6}~Z_{\odot}$ by the SN remnant and will likely host normal Pop~II stars. Most minihalos have collapsed to high enough densities by the time star formation ensues and thus remain pristine.}
\end{center}
\end{figure}

\begin{figure}
\begin{center}
\resizebox{8cm}{20cm}
{\unitlength1cm
\begin{picture}(8,20)
\put(0.0,2.0){\includegraphics[width=8.0cm,height=18.0cm]{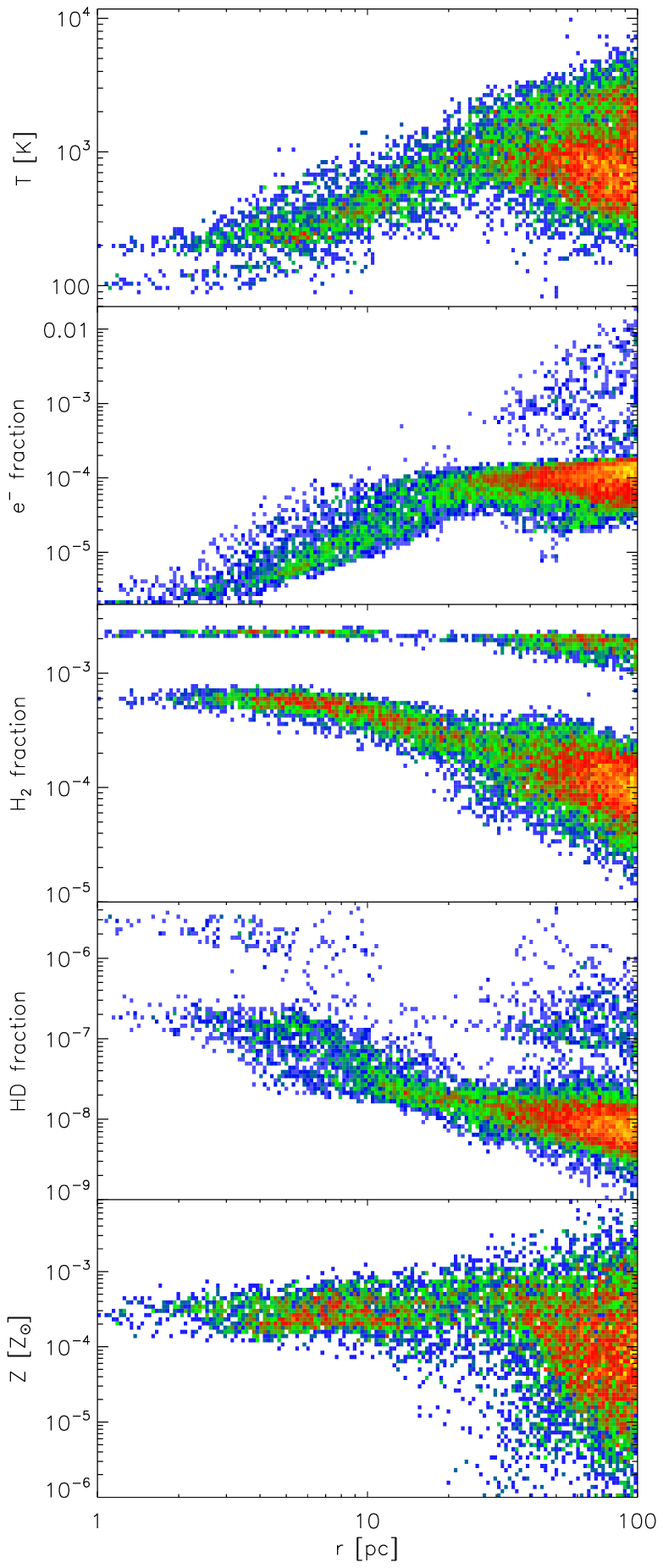}}
\put(0.0,0.0){\includegraphics[width=8.0cm,height=2.0cm]{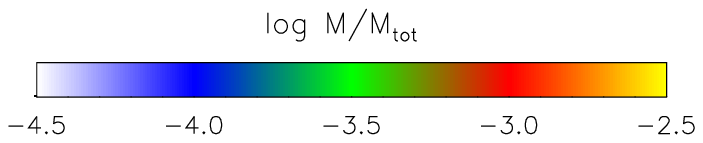}}
\end{picture}}
\caption{Radial distribution of gas in a minihalo that has been partially shock-heated by an ionization front and subsequently disrupted by the SN remnant (Sim~A), color-coded according to its mass fraction. The temperature panel shows that the core has fragmented into two distinct clumps that are both enriched to $Z\simeq 3\times 10^{-4}~Z_{\odot}$: one in which the photoionization has led to enhanced molecule abundances, with H$_{2}$ and HD fractions $\ga 10^{-3}$ and $\ga 10^{-8}$, respectively, allowing the gas to cool to $100~{\rm K}$, and another where the gas has shielded itself from radiation and has only cooled to $\sim 200~{\rm K}$. The CMB temperature at these redshifts is $\simeq 50~{\rm K}$. Metal fine-structure cooling does not become important, although dust-induced cooling may set in at higher densities than we can resolve here.}
\end{center}
\end{figure}

\begin{figure}
\begin{center}
\resizebox{8cm}{21cm}
{\unitlength1cm
\begin{picture}(8,21)
\put(0.0,2.0){\includegraphics[width=8.0cm,height=19.0cm]{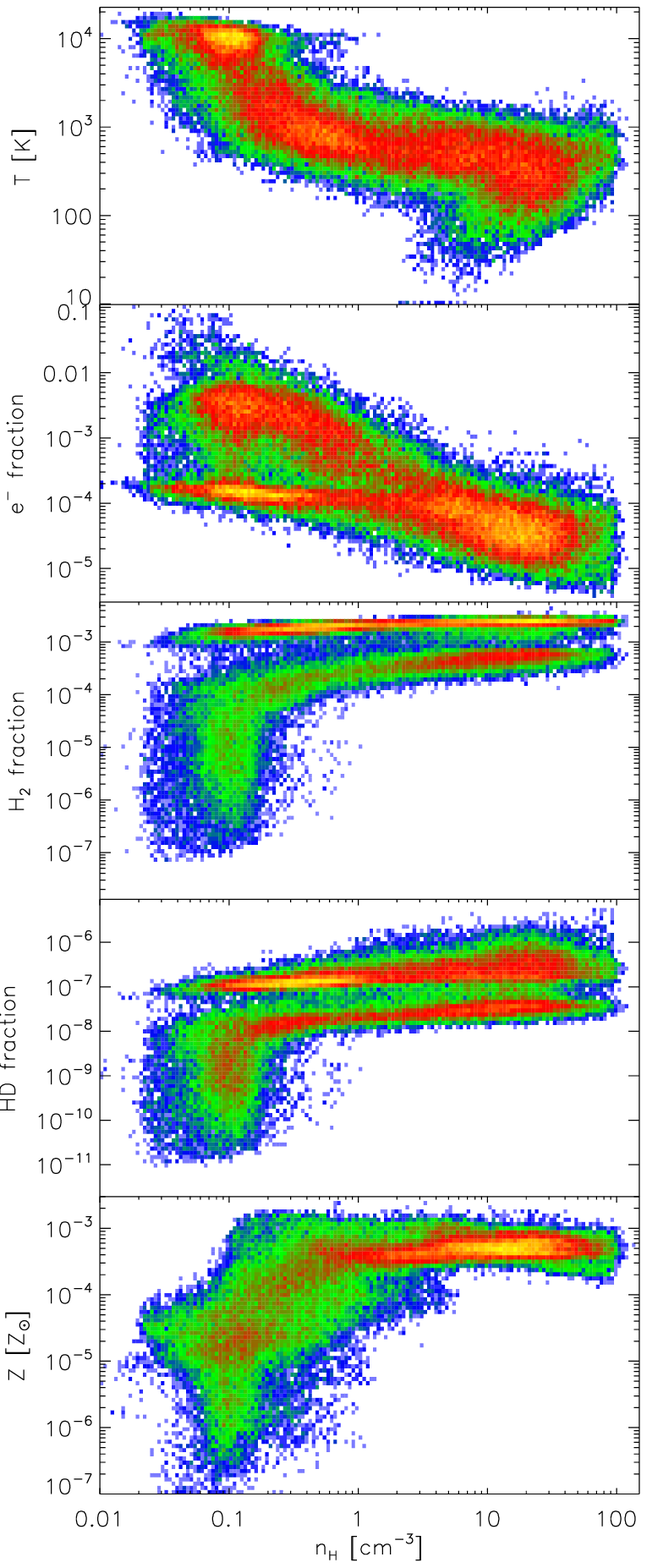}}
\put(0.0,0.0){\includegraphics[width=8.0cm,height=2.0cm]{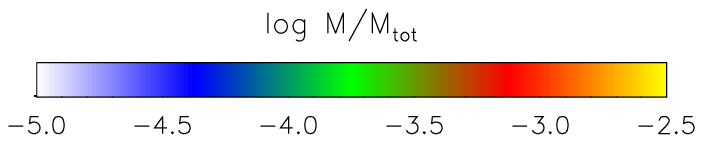}}
\end{picture}}
\caption{Distribution of gas in density and temperature space within the virial radius of the galaxy (Sim~A), color-coded according to its mass fraction. The dense gas at the center of the galaxy is enriched to $Z\sim 10^{-3}~Z_{\odot}$ and will likely form of a stellar cluster with a more normal IMF. Similar to the minihalo case shown in Figure~8, the photoheated gas has formed enhanced molecule abundances and will likely cool to the temperature of the CMB at somewhat higher densities.}
\end{center}
\end{figure}

\subsection{Star formation in minihalos}

Concomitant to the assembly of the galaxy, nearby minihalos undergo runaway collapse and form stars. Depending on distance and central density, their collapse could be unaffected, delayed, or completely inhibited in the presence of an ionization front or the SN remnant. Recent detailed radiation-hydrodynamics simulations performed by \citet{whalen08a,whalen08b,whm10} have revealed that for a large region in parameter space most minihalos remain unaffected. However, in distinct cases the ionization front or the SN shock can disrupt the core of the halo, heating and ionizing the gas and possibly leading to metal enrichment. This in turn may trigger the formation of less massive so-called Pop~III.2 stars \citep{oshea05,jb06,yoh07}, or normal Pop~II stars if enough metals are brought to the center of the halo. The amount of fragmentation sensitively depends on the initial conditions of the collapse \citep{jappsen09b}.

How does this compare with our results? In Figure~7, we show all star-forming minihalos in the high resolution region as a function of redshift and distance from the center of the galaxy. Crosses denote minihalos that remain largely unaffected by feedback effects and will likely form Pop~III.1 stars. Minihalos that have been severely disturbed by ionizing radiation from a nearby star show elevated electron and H$_{2}$ fractions and are denoted by star symbols. These halos also form significant amounts of HD and will most likely cool to the temperature of the CMB before becoming Jeans-unstable, following the canonical pathway of Pop~III.2 stars. Finally, triangles denote minihalos that have been disrupted by the SN remnant and are enriched to $Z>10^{-6}~Z_{\odot}$. Two of these halos are even enriched to $Z>10^{-3.5}~Z_{\odot}$, and will almost certainly form Pop~II stars.

In Figure~8, we show a particularly interesting case that elucidates the importance of timing in regulating the strength of feedback. Here, a minihalo has been partially photo-heated by a nearby star to $10^{4}~{\rm K}$, and subsequently enriched with metals. In the process of its collapse, the core has fragmented into two distinct objects: one in which the molecule fraction has been significantly enhanced due to the photoionization, allowing the gas to cool to $\simeq 100~{\rm K}$, and another in which the cooling is unaltered from the standard minihalo pathway. In both cases, the metallicity is not high enough for metal fine-structure cooling to become important, although at higher densities dust-induced cooling may take over and lead to the formation of a small cluster of Pop~II stars. This case suggests that the formation of low-mass stars need not await the formation of second-generation halos with $M_{\rm vir}\ga 10^{8}~M_{\odot}$.

\subsection{Star formation in the nascent galaxy}

What are the properties of the gas within the newly virialized galaxy? From Figures~5 and 6, it appears that $\sim 10^5~M_{\odot}$ of cold, dense gas at its center have been enriched to $Z\sim 10^{-3}~Z_{\odot}$. This is confirmed by Figure~9, where we show its distribution in density and temperature space within the virial radius of the galaxy. It has become highly enriched by accretion from the surrounding IGM and is in a state of collapse. We also find an interesting effect related to photoheating: previously ionized gas forms elevated H$_{2}$ and HD abundances and coexists with unheated gas. Similar to the minihalo case discussed in Figure~8, the gas might therefore later fragment and form individual clumps. We note that in the density range resolved here, cooling is dominated by primordial molecules instead of metal fine-structure cooling \citep{jappsen07,jappsen09a}. At somewhat higher densities, we expect that the gas will cool to the temperature of the CMB before the equation of state hardens again. However, a more detailed investigation of the subsequent fragmentation must await high-resolution simulations that resimulate the central few kpc of the galaxy.

\section{Summary and conclusions}

We have performed a set of highly resolved SPH simulations that allow us to investigate the enrichment of the IGM by a PISN exploding in a high-redshift minihalo. We have incorporated a substantially higher degree of realism compared to our previous work in the form of radiation feedback, explicit chemical mixing, and metal line cooling. In particular, we have employed a physically motivated model for the mixing of metals between individual SPH particles based on diffusion \citep{greif09a}. We have followed the distribution of metals as they are ejected into the IGM and then recollapse into the larger, $M_{\rm vir}\sim 10^{8}~M_{\odot}$ potential well of a ``first galaxy'' assembling at $z\simeq 10$. We have performed simulations with and without radiative feedback to assess the effects of photoheating on the assembly of the galaxy and the distribution of metals.

The heavy elements ejected by the SN are initially distributed into the IGM by the bulk motion of the SN remnant, until it stalls and mixing is instead facilitated by turbulent motions on smaller scales. These are induced by the dynamics of the underlying DM, and, to a smaller degree, by the additional photoheating. A clear correlation between metallicity and gas overdensity is established, where the densest regions consisting of existing minihalos remain largely pristine, while voids around the SN progenitor become highly enriched. This correlation breaks down once the potential well of the galaxy assembles and metal-rich gas residing in the IGM recollapses to high densities. The metallicity at the center of the galaxy then grows to $Z\sim 10^{-3}~Z_{\odot}$, likely resulting in the formation of a stellar cluster with a more normal IMF \citep{cgk08}. Although the influence of photoheating on the mixing of the gas is limited, an interesting side effect is that it can expel enriched gas out of the potential well of the galaxy into the IGM. Finally, the mechanical feedback exerted by the photoheating and the propagation of the SN remnant is usually quite small, although in some cases nearby minihalos are disrupted and enriched, possibly leading to the formation of Pop~III.2 and Pop~II stars.

One of the most promising observational tools to understand the nature of these objects relies on the elemental composition of second-generation stars. Previous searches for PISN signatures have targeted the most metal-poor stars in our Galaxy, with an iron abundance as low as $[{\rm Fe}/{\rm H}]\sim 10^{-5}$ \citep[e.g.,][]{bc05,frebel05}. Unfortunately, the spectra of these stars do not reveal the strong odd--even effect indicative of PISNe, implying that they were either not very common, or that existing surveys did not target a suitable sample of stars. Indeed, our simulations show that a single PISN is sufficient to enrich a gas cloud quite uniformly to $Z\sim 10^{-3}~Z_{\odot}$, implying that its signature may be buried in stars with a significantly higher metallicity \citep{kjb08}. The ideal search strategy may therefore have to be modified to concentrate on stars with $[{\rm Fe}/{\rm H}]\sim 10^{-3}$. Recent surveys of extragalactic globular clusters lend support to this idea, since the oldest of these have been found to contain large $[\alpha /{\rm Fe}]$ ratios, which is indicative of enrichment by PISNe \citep{pkg06}.

The standard picture that minihalos only form Pop~III stars, or remain sterile before being disrupted during subsequent merging \citep[e.g.,][]{hh03}, may also have to be modified. Whereas that is indeed the case for the majority of minihalos in our simulation, there exists a small number of halos that do contain dense, metal-enriched gas. Although we do not follow its subsequent evolution, it is intriguing to speculate whether these clouds might fragment to form the first globular clusters, or at least slightly less massive precursors \citep[e.g.,][]{bc02,boley09}. Soon, it should become possible to study the further fate of these second-generation star forming clouds with the required extremely high resolution by resimulating the central halo gas, and by including dust cooling and opacity effects at high density. Similar simulations will be able to address the formation of the first star clusters inside the first galaxies, starting from the initial conditions determined here. The goal of understanding galaxy formation from first principles, at least in the relatively simple case of dwarf-sized systems at the beginning of hierarchical structure formation, might finally have come into reach.

\acknowledgements{T.H.G thanks Simon White for many stimulating discussions, as well as Klaus Dolag for technical support. V.B. acknowledges support from NSF grant AST-0708795 and NASA ATFP grant NNX08AL43G. The simulations were carried out at the Texas Advanced Computing Center (TACC), under TeraGrid allocation TG-AST090003. S.C.O.G and R.S.K. acknowledge financial support from the German {\em Bundesministerium f\"{u}r Bildung und Forschung} via the ASTRONET project STAR FORMAT (grant 05A09VHA) and from the {\em Deutsche Forschungsgemeinschaft} (DFG) under grants no. KL 1358/1, KL 1358/4, KL 1359/5, KL 1358/10, and KL 1358/11. S.C.O.G. and R.S.K. furthermore acknowledge support from subsidies from a Frontier grant of Heidelberg University sponsored by the German Excellence Initiative and from the {\em Landesstiftung Baden-W\"{u}rttemberg} via their program International Collaboration II (grant P-LS-SPII/18). In addition, R.S.K. thanks the KIPAC at Stanford University and the Department of Astronomy and Astrophysics at the University of California at Santa Cruz for their warm hospitality during a sabbatical stay in spring 2010.}

\bibliographystyle{apj}

\end{document}